\documentclass[preprint2,numberedappendix,appendixfloats]{aastex6}
\usepackage{amstext}
\usepackage{amsmath}
\usepackage{amssymb}
\usepackage{apjfonts}
\usepackage{verbatim}
\begin{document}

\newcommand{\beq}{\begin{equation}}
\newcommand{\eeq}{\end{equation}}
\newcommand{\beqar}{\begin{eqnarray}}
\newcommand{\eeqar}{\end{eqnarray}}

\newcommand{\ebv}{E(B-V)}

\title{Lessons from the Onset of a Common Envelope Episode:  the Remarkable M31 2015 Luminous Red Nova Outburst}

\author{Morgan MacLeod\altaffilmark{1}\altaffilmark{2}}

\author{Phillip Macias\altaffilmark{3}}

\author{Enrico Ramirez-Ruiz\altaffilmark{3}}

\author{Jonathan Grindlay\altaffilmark{4}}

\author{Aldo Batta\altaffilmark{3}}
\and
\author{Gabriela Montes\altaffilmark{3}}

\altaffiltext{1}{School of Natural Sciences, Institute for Advanced Study, 1 Einstein Drive, Princeton, New Jersey 08540, USA}

\altaffiltext{2}{NASA Einstein Fellow}

\altaffiltext{3}{Department of Astronomy \& Astrophysics, University of California, Santa Cruz, CA 95064, USA}

\altaffiltext{4}{Harvard-Smithsonian Center for Astrophysics, The Institute for Theory and Computation, 60 Garden Street, Cambridge, MA 02138, USA}

\begin{abstract} 
This paper investigates the recent stellar merger transient M31LRN 2015 in the Andromeda galaxy. 
We analyze published optical photometry and spectroscopy along with a Hubble Space Telescope detection of the color and magnitude of the pre-outburst source. 
The transient outburst is consistent with dynamically driven ejecta at the onset of a common envelope episode, which eventually leads to the complete merger of a binary system. The light curve appears to contain two components: first $\sim10^{-2} M_\odot$ of fast ejecta driven by shocks at the onset of common envelope, and later, $\sim0.3 M_\odot$ of further ejecta as the secondary becomes more deeply engulfed within the primary. 
Just prior to merger, we find that the primary star is a $3-5.5 M_\odot$ sub-giant branch primary star with radius of $30-40R_\odot$. Its position in the color-magnitude diagram shows that it is growing in radius, consistent with a picture where it engulfs its companion. 
By matching the properties of the primary star to the transient outburst, we show that the optical transient lasts less than ten orbits of the original binary, which had a pre-merger period of $\sim 10$ days. 
We consider the possible orbital dynamics leading up to the merger, and argue that if the system merged due to the Darwin tidal instability it implies a lower mass main sequence companion of $0.1-0.6M_\odot$.  This analysis represents a promising step toward a more detailed understanding of flows in common envelope episodes through direct observational constraints.
\end{abstract}

\section{Introduction}

Many stars, especially those more massive than the sun, live in binary or multiple systems \citep[][]{2013ARA&A..51..269D,2016arXiv160605347M}. 
As stars evolve off the main sequence and their radii grow significantly, binary systems that were non-interacting can evolve into contact. 
Among triple and higher-order multiple systems, secular interactions can drive pairs of objects into eccentric orbits that lead to their direct impact \citep[e.g.][]{1962AJ.....67..591K,2011ApJ...741...82T}.
Through these interaction channels, stellar multiplicity plays a key role in shaping the evolution of stellar populations \citep[e.g.][]{2012Sci...337..444S,2013ApJ...764..166D}. 

The outcome of a stellar interaction is shaped by the orbital dynamics, relative masses and evolutionary states of the component stars.  Tidally synchronized close binaries which come into contact on the main sequence can form peanut-shaped, yet stable, overcontact systems known as W UMa stars \citep[e.g.][]{1976ApJ...209..536S,1976ApJ...209..829W,1977MNRAS.179..359R,1995ApJ...444L..41R}. Other interacting pairs, especially those with a more-massive evolved-star donor, become unstable at or before the onset of mass transfer. 
This instability can arise from runaway mass exchange (when a star expands more quickly than its Roche lobe on losing mass) or from destabilizing angular momentum exchange between the objects in the binary system  (as in the \citet{1879RSPS...29..168D} tidal instability). In either of these cases the component objects are driven toward  merger. 

The runaway merger of two stars leads to either a remnant composed of the bulk of the mass of the pair, or to the formation of a new, tighter binary. In the case of a remnant binary, the new pair formed by the secondary star along with the core of the primary must have injected enough energy into the gaseous \emph{common envelope} donated by the primary star to clear its surroundings and stabilize as a system transformed by a phase of orbital inspiral \citep[][]{1976IAUS...73...75P,1979A&A....78..167M,1993PASP..105.1373I,2000ARA&A..38..113T,2010NewAR..54...65T,2013A&ARv..21...59I}. 
These phases of orbital transformation are key in shaping populations of compact binaries which interact or merge \citep{2007PhR...442...75K,2014LRR....17....3P} -- producing some of the most dramatic electromagnetic and gravitational transients when they do \citep{2016ApJ...818L..22A}.  Yet the hydrodynamics, overall efficiency, and division of systems that merge and those that eject their envelopes following these phases of unstable binary interaction remain poorly-understood subjects of intense scrutiny \citep{2013A&ARv..21...59I}. 

In order to improve our understanding of common envelope episodes, we can rely on before-and-after comparisons of stellar populations or attempt to catch common envelope events and stellar mergers in action \citep[e.g.][]{2013Sci...339..433I,2013A&ARv..21...59I}. 
The onset of common envelope episodes and binary mergers proceeds similarly -- particularly if the system is composed of one evolved star and a more compact companion. A  distinction arises later, following a phase of orbital inspiral, when the bulk of the envelope material either is (or is not) driven off \citep[e.g.][]{1993PASP..105.1373I,2011MNRAS.411.2277D,2008ApJ...672L..41R,2012ApJ...746...74R}. 
At the onset of interaction between two stars, a small portion of mass will be ejected, powering an optical or infrared transient as it expands and becomes transparent \citep{2006MNRAS.373..733S,2012MNRAS.425.2778M,2013Sci...339..433I}. The detection and detailed study of this category of stellar merger and common envelope transient therefore offers direct constraints on the conditions and flow properties at the onset of this highly uncertain phase of binary evolution. 

An emergent class of transients --  luminous red novae (LRN) -- have come to be associated with stellar mergers through detailed study of a few key events. M31 RV was one of the first red transients to be identified, in 1988, but its lightcurve is only captured during the decline \citep{1990ApJ...353L..35M,1992PASP..104..179B,2004A&A...418..869B,2006AJ....131..984B,2011ApJ...737...17B}. A galactic transient, V838 Mon, illuminated its surroundings with a spectacular light echo imaged by the Hubble Space Telescope (HST) from 2002-2008 following its outburst \citep{2002IAUC.7892....2B,2002A&A...389L..51M,2002MNRAS.336L..43K,2003Natur.422..405B,2005A&A...434.1107M,2008AJ....135..605S,2010ApJ...717..795A,2014A&A...569L...3C,2014Msngr.158...35M}.  The object V1309 Sco, another galactic transient, proved critical in establishing the origins of this class of events \citep{2010A&A...516A.108M,2013MNRAS.431L..33N}. A multi-year time series of photometric data taken by OGLE \citep{1998AJ....115.1135R,2008AcA....58...69U} revealed an eclipsing binary with decreasing orbital period prior to the outburst \citep{2011A&A...528A.114T,2011A&A...531A..18S}. Following the event, a single object remained \citep{2011A&A...528A.114T,2013MNRAS.431L..33N,2015A&A...580A..34K}. 

Other recently-discovered transients populate a similar (or slightly longer duration and more luminous) phase space in a luminosity-timescale diagram \citep{2012PASA...29..482K}. These include objects like NGC 300-OT \citep{2009ApJ...695L.154B,2009ApJ...699.1850B}, PTF 10fqs \citep{2011ApJ...730..134K}, M85-OT \citep{2007Natur.447..458K,2007ApJ...659.1536R,2008ApJ...674..447O}, and supernova 2008S \citep{2009ApJ...697L..49S}, and are generally categorized as intermediate-luminosity optical transients with a massive star-outburst (rather than binary) origin \citep[e.g.][]{2009ApJ...705.1364T,2011ApJ...741...37K,2016MNRAS.458..950S}.

This paper focuses on a new, and particularly remarkable, addition to the category of stellar merger transients: M31LRN 2015. Discovered in January 2015 by the MASTER network\footnote{http://observ.pereplet.ru/}, the transient resides in the Andromeda galaxy.  It was discovered about 8 days prior to peak brightness, and both \citet{2015ApJ...805L..18W} and \citet{2015A&A...578L..10K} have published multicolor photometry of the outburst light curve along with spectra. What makes this event particularly useful is the existence of multiband pre-outburst imaging of the source in the years prior to the transient \citep{2015ATel.7173....1D,2015ApJ...805L..18W}.  Since the association with M31 (and thus the distance) is known, the pre-outburst color and magnitude allow us to compare the physical properties of the primary star in the presumed binary system to the merger-driven outburst it produces.  Uncertainties in the progenitor mass and color -- in the case of V838 Mon \citep{2005A&A...434.1107M,2005A&A...441.1099T} -- or in it's distance -- in the case of V1309 Sco \citep{2011A&A...528A.114T} -- have hindered such an analysis of previous well-studied transients.  Our analysis of this transient is complimented by a concurrent study of another extragalactic transient presumed to be driven by a stellar merger in M101 \citep{2016arXiv160708248B}.

Our analysis of this source proceeds as follows. In Section \ref{sec:outburst}, we  focus on the transient outburst, and estimate some physical properties of the ejecta from the photometric and spectroscopic data published by \citet{2015ApJ...805L..18W} and \citet{2015A&A...578L..10K}. In Section \ref{sec:progenitor}, we consider the pre-outburst detection by HST, and compare to stellar tracks to assess the primary-star's mass, radius, and internal structure from its position in the color-magnitude diagram. In Section \ref{sec:comb_req}, we consider the properties of the transient that emerge when we consider both the outburst and the pre-merger source together, and propose a model for interpreting the data. Section \ref{sec:mergers} extends our discussion to consider the pathways that a binary may take to the onset of a common envelope phase. In the context of these arguments, Section \ref{sec:DI} proposes that M31 LRN 2015 might have originated from a Darwin-unstable pair of stars, and studies the implications of this proposal.  
In Section \ref{sec:discussion}, we discuss M31LRN 2015 in the context of other similar transients and conclude.

\section{Outburst}\label{sec:outburst}

\subsection{Summary of Observations}
\begin{figure}[tbp]
\begin{center}
\includegraphics[width=0.48\textwidth]{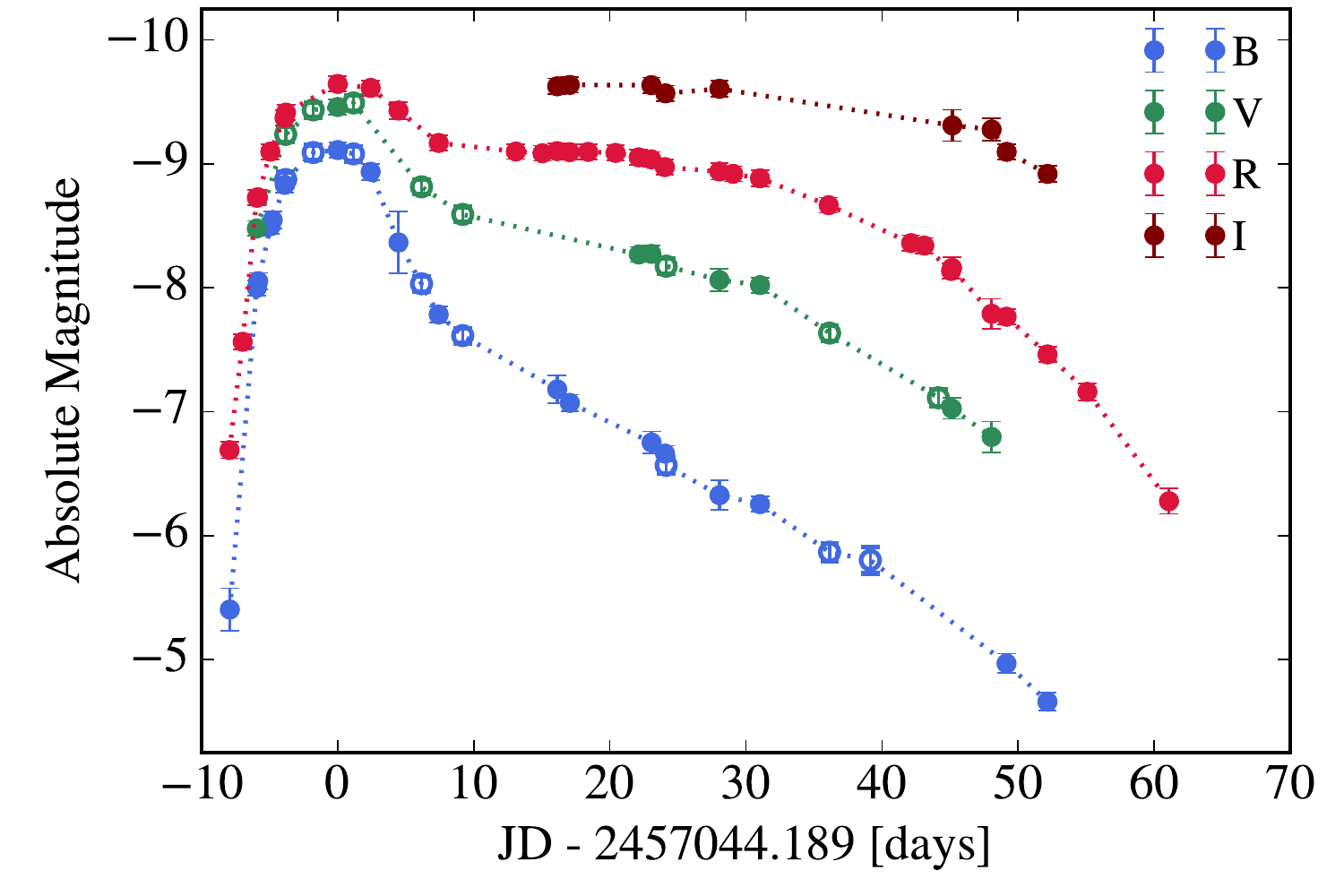}
\caption{Absolute light curve of the optical outburst of M31LRN 2015. Data from \citet{2015ApJ...805L..18W} (open points) and \citet{2015A&A...578L..10K} (closed points). This figure is constructed assuming $\ebv=0.15$~mag and error bars show measurement and distance errors while ignoring reddening error. The similarity of this lightcurve to other LRN transients like V838 Mon marked M31LRN 2015 as a stellar merger.  }
\label{fig:lc}
\end{center}
\end{figure}

M31LRN 2015 (MASTER OT J004207.99+405501.1) was discovered in January 2015 \citep{2015ApJ...805L..18W}, and both \citet{2015ApJ...805L..18W} and \citet{2015A&A...578L..10K} have since published photometry and spectroscopy of the outburst. The source is located in M31 \citep{2015ApJ...805L..18W}. While the distance is therefore well known, the reddening to the source has been estimated at $\ebv=0.12 \pm 0.06$~mag \citep{2015ApJ...805L..18W} or $\ebv=0.35\pm0.1$~mag  \citep{2015A&A...578L..10K} and thus remains a source of uncertainty. 
In Figure \ref{fig:lc}, we show the outburst absolute light curve using photometric data from both \citet{2015ApJ...805L..18W} and \citet{2015A&A...578L..10K} under the assumption of $\ebv=0.15$~mag. 
The transient peaks at $M_V \approx -9.5$, and reddens progressively during its decline. The elapsed time from discovery to peak is approximately 8~d, and the optically-bright portion of the light curve persists for $\sim 50$~d more. 

These photometric data have been supplemented by spectroscopic observations reported both by \citet{2015ApJ...805L..18W} and \citet{2015A&A...578L..10K}. \citet{2015ApJ...805L..18W}'s data has a spectral resolution of 18~\AA, or $R\approx360$  at 6500~{\AA} (821~km~s$^{-1}$).  A spectrum taken 3.2 days after the discovery of the transient and 4.7 days prior to peak shows a strong H$\alpha$ emission feature with an uncorrected FWHM of $900\pm100$~km~s$^{-1}$ \citep{2015ApJ...805L..18W}.  When corrected to instrumental resolution (by subtracting the instrumental resolution in quadrature) this suggests an intrinsic FWHM that may be associated with the expanding ejecta of 
\beq\label{williams_v}
v_{\rm ej} = 370\pm240{\rm~km~s}^{-1}.
\eeq
Over time, the  H$\alpha$ emission fades and Na I D and Ba II absorption features appear. 
\citet{2015A&A...578L..10K} also report H$\alpha$ emission in early spectra, but they do not report an emission line FWHM. 
In an $R=1000$ spectrum covering 4000-5000~{\AA} taken 6 days before peak, \citet{2015A&A...578L..10K} report a host of absorption lines and a similarity to a stellar F5I spectrum. 

\subsection{Modeled Properties}

\begin{figure*}[tbp]
\begin{center}
\includegraphics[width=0.85\textwidth]{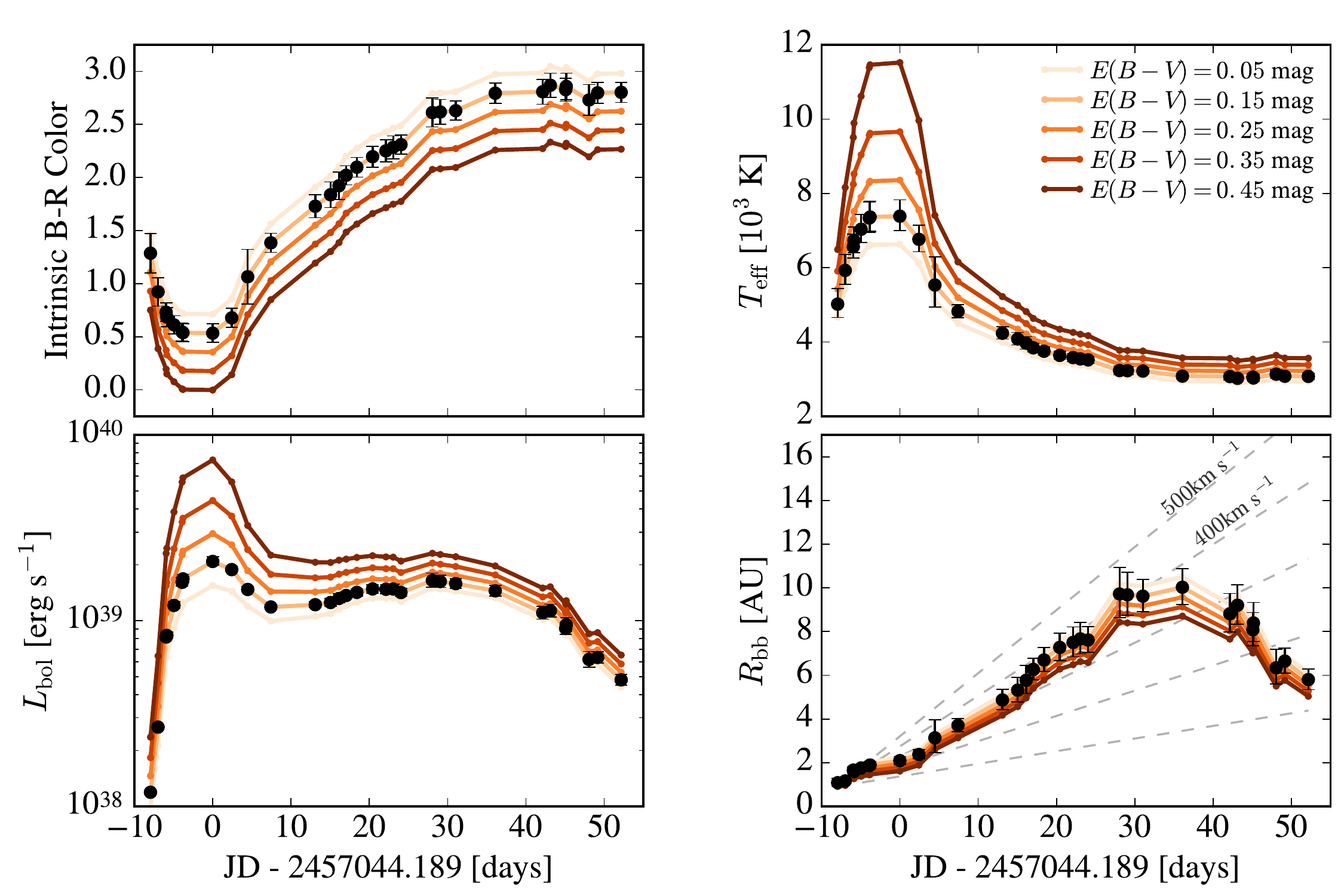}
\caption{Photometric properties of the outburst modeled as blackbody emission. Overset lines in each panel represent different assumed reddening, $\ebv$. Points and error bars are plotted for  $\ebv=0.15$~mag with the assumption of {\it no} error in the reddening vector itself. 
The transient first becomes bluer with peak temperature of $10^4$~K at the time of lightcurve peak,  then progressively redder. The bolometric lightcurve shows two components 1) a rapid rise to peak lasting $\sim 8$~d, followed by a brief fading of similar duration and 2) a longer lived plateau extending from 10 to 50~d after the time of peak. 
During this time the photosphere radius increases from $\sim 2$~AU around peak to a maximum of $\sim 10$~AU approximately 35~d after peak before beginning to recede. 
In the background of this panel we mark lines of constant expansion velocity of 100-500~km~s$^{-1}$. 
}
\label{fig:phot}
\end{center}
\end{figure*}

In this section we use the photometry from \citet{2015ApJ...805L..18W} and \citet{2015A&A...578L..10K} shown in Figure \ref{fig:lc} to derive some physical properties of the transient outburst under the assumptions that the ejection of mass in a stellar merger powers the transient and of a blackbody spectral energy distribution.  

Because of the uncertain source reddening, we apply corrections for $\ebv$ between 0.05~mag and 0.45~mag (in increments of 0.1~mag), spanning the full range of estimates by  \citet{2015ApJ...805L..18W} and \citet{2015A&A...578L..10K}. 
Figure \ref{fig:phot} shows different values of the reddening to the source as different line colors. We plot the data points (in black) only for one value of the reddening, $\ebv=0.15$~mag.  The plotted error bars represent the propagation of photometric measurement errors with the assumption of no reddening uncertainty. 

We use the $B-R$ color (upper left panel) to estimate the effective temperature assuming that the spectral energy distribution is a blackbody (upper right panel).  With the effective temperature, we derive the $R$-band bolometric correction, and apply that to find $M_{\rm bol}$ and $L_{\rm bol}$ (lower left panel).  
Finally, assuming $L_{\rm bol}=4\pi R_{\rm bb}^2 \sigma_{\rm b} T_{\rm eff}^4$, we derive the radius of the blackbody photosphere (lower right panel) . 
The effective temperature starts around 6000~K at the time of the first observations, and peaks near the time of maximum brightness of the transient, with maximum temperatures of $\sim 10^4$~K (dependent on reddening). Over the following $\sim40$~d, the transient's photosphere becomes cooler, with inferred temperatures falling to the 3000-4000~K range.

During the time mapped by the observations, the inferred photosphere radius, $R_{\rm bb}$, first expands, then recedes. Near the time of peak the photosphere radius is approximately 2~AU. It reaches a maximum of approximately 10~AU about 35~d after the lightcurve peak. 
The velocity of this expansion and the maximum photosphere radius reached are a weak function of the assumed reddening. Lines of constant expansion velocity are shown in the background of Figure \ref{fig:phot}, and the typical expansion velocity is $\sim 400$~km~s$^{-1}$.

The transient's bolometric light curve exhibits a two-part structure consisting of an early rise to peak and subsequent decay (times of approximately -10 to +10~days relative to the peak time in Figure \ref{fig:phot}) followed by a longer plateau of approximately constant bolometric luminosity. 
On the basis of this observation, we will refer to two portions of the lightcurve as the {\it peak} (times of $\pm 10$~d relative to peak at JD 2457044.189) and {\it plateau} (times of 10 to 50~d after peak).\footnote{We note that the late time fade in optically-inferred bolometric luminosity might not reflect the full energetics if there is a significant dust reprocessing peak in the spectral energy distribution, as was observed at late times in V1309 Sco by \citet{2016A&A...592A.134T} and in NGC 4490 OT by \citet{2016MNRAS.458..950S}.}

\begin{figure}[tbp]
\begin{center}
\includegraphics[width=0.45\textwidth]{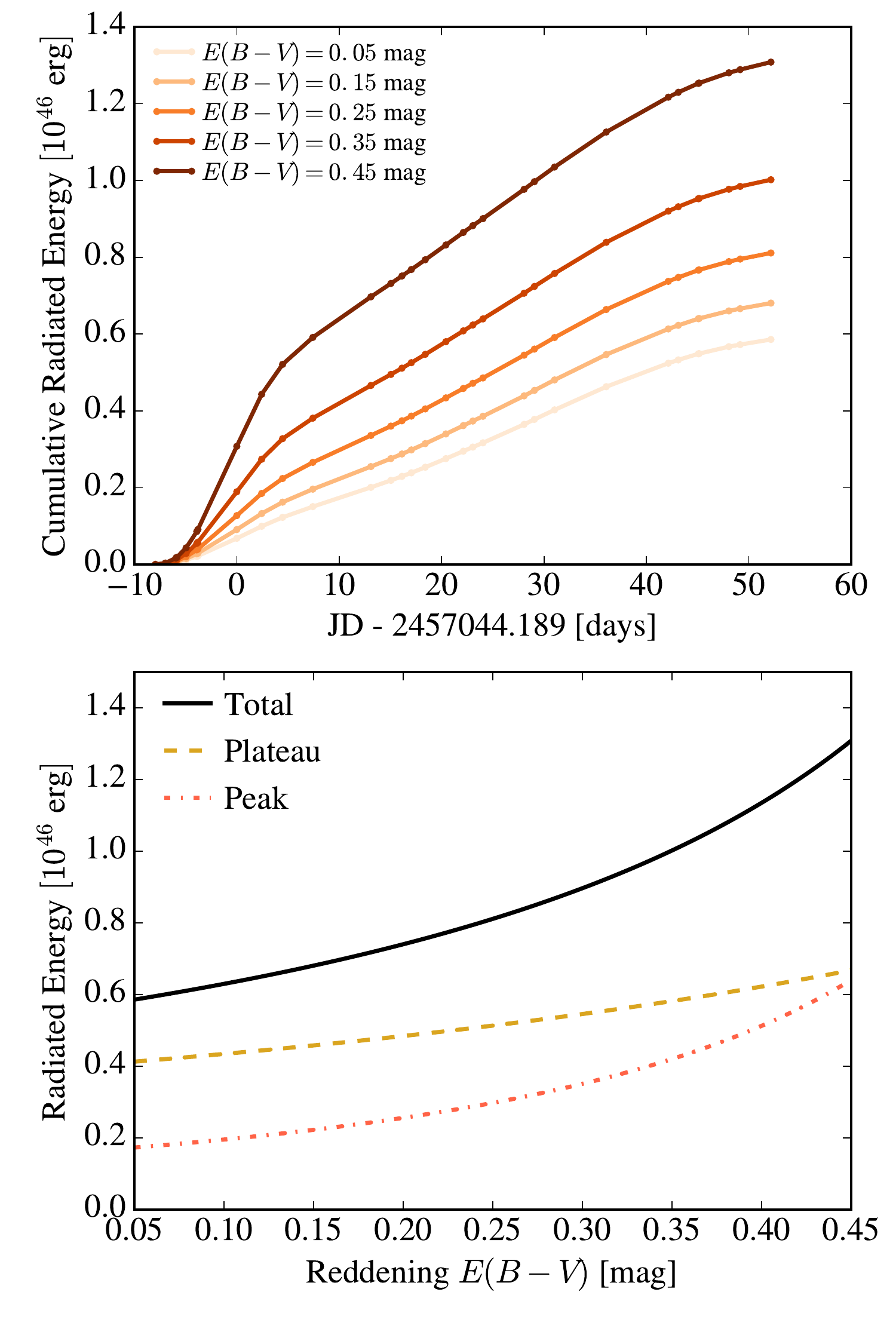}
\caption{The radiated bolometric energy inferred from the optical light curve of M31LRN 2015. The upper panel shows the cumulative radiated energy throughout the outburst for the same reddening assumptions as in Figure \ref{fig:phot}. The lower panel plots the integral quantities as a function of reddening to the source. The contributions from the peak and plateau sub-components are computed from the ${\pm 10}$~d and the 10-50~d portions of the light curve shown in Figure \ref{fig:phot}. The radiated energy is dominated by the plateau in all cases, with the peak contributing roughly 30-50\% of the total. Regardless of the reddening, the total radiated energy is within a factor of two of $10^{46}$~erg.  }
\label{fig:erad}
\end{center}
\end{figure}

Figure \ref{fig:erad} explores the radiated energy during the outburst. The upper panel shows the cumulative radiated energy. This rises steeply in the peak portion of the lightcurve, and more shallowly after peak. The relative slopes are determined by the reddening. When the reddening is high, the peak is most accentuated, as is also the case in the lower-left panel of Figure \ref{fig:phot}. 
The relative contributions of the peak and plateau to the total integrated radiated energy ($\sim 10^{46}$ erg) are also shown in Figure \ref{fig:erad}.  
The plateau always dominates the energy release, with the peak contributing 30-50\% of the radiated energy, or an energy of $2-4\times10^{45}$~erg.

\subsection{ Estimates of Ejecta Mass and Outburst Energetics}

Although the properties of the merger-driven outflow are likely to be complex, we can still draw some important conclusions from the physical properties inferred from the light curve, especially when paired with  the spectroscopic data. 
In this section we divide our focus based on the two key phases of the bolometric light curve as visible in the lower left panel of Figure \ref{fig:phot}: the peak (${\pm 10}$~d) and the plateau (10-50~d). 

To begin, we estimate the minimum mass needed to produce an opaque shell at the observed blackbody radius $R_{\rm bb}$ by assuming that the obscuring mass is spread over a sphere of radius $R_{\rm bb}$ with constant density, $\rho$. We then estimate the gas mass $\Delta m_{\rm obsc}$ needed to produce an optical depth,  $\tau = \rho \kappa R_{\rm bb} = \Delta m_{\rm obsc} \kappa / (4/3 \pi R_{\rm bb}^2)$, of order unity,
\beq
\Delta m_{\rm obsc} \approx 1.4 \times 10^{-6} M_\odot \left( \frac{R_{\rm bb}  }{\rm AU} \right)^2 \left( \frac{\kappa  }{0.34 {\rm ~cm^2~g^{-1}}} \right)^{-1},
\eeq 
where we take a fiducial opacity based on electron scattering in ionized solar-composition material. We note that if the material is only partially ionized, the opacity could be lower by up to several orders of magnitude.  This estimate nonetheless shows that even a tiny ejecta mass can be responsible for the growth in the emitting surface.

\subsubsection{Early Peak}\label{sec:early_peak_scalings}
Here we  focus on the  early portion of the light curve in order to derive some additional constraints.  The minimum obscuring mass (estimated above) is sufficient to explain the growth of $R_{\rm bb}$. But it cannot explain the rise and fall timescales of the light curve, which require that heat does not diffuse out of the expanding gas instantaneously. When the photon mean free path is of order the size of the object, the heat diffusion time is of order the light crossing time -- just 8 minutes for 1~AU. 

Instead, we use the behavior of an expanding shell of gaseous ejecta with diffusion but no other heating or cooling, following \citet[][chapter 4.8]{2001thas.book.....P}. In this model, the gas is heated to the virial temperature during its interaction with the secondary. It expands, becomes more transparent, and this heat diffuses toward the photosphere, generating a shift to higher luminosity and effective temperature. H$\alpha$ emission lines during this phase indicate that a recombination front lies outside the photosphere and that the gas (at the photosphere location) is likely ionized during this portion of the lightcurve. As a result, we adopt an opacity $\kappa = 0.34$~cm$^2$~g$^{-1}$ in our subsequent analysis. The estimates of such a model are certainly approximate, but provide useful insight into the properties of the early ejecta in this transient. 

The first law of thermodynamics implies that the light curve will peak at $\tau_{\rm peak} = \sqrt{2 \tau_d \tau_h}$, where $ \tau_d $ is the diffusion time of photons through the gas at the initial radius $ \tau_d = f \kappa \Delta m_{\rm ej,peak} /( c r_0)$. The constant is $f\approx0.07$ for a spherical shell \citep{2001thas.book.....P}.   $\tau_h = r_0/v_{\rm ej}$ is the hydrodynamic timescale based on the initial radius $r_0$ and velocity $v_{\rm ej}$. 
The time of peak is  
\beq\label{tpeak}
\tau_{\rm peak} \approx 10.7 {\rm \ d} \left(  \frac{ \Delta m_{\rm ej,peak}}{0.01 M_\odot} \right)^{1/2} \left(  \frac{v_{\rm ej}}{370 \ {\rm km \ s}^{-1}} \right)^{-1/2},
\eeq
with $\kappa = 0.34$~cm$^2$~g$^{-1}$. 
We've substituted a characteristic velocity of the outflowing material based on the corrected H$\alpha$ FWHM velocity, $v_{\rm ej}=370$~km~s$^{-1}$, equation \eqref{williams_v}, measured by \citet{2015ApJ...805L..18W} approximately $4.7$~d before the lightcurve peak.

We can re-express the diffusion argument above to solve for the ejected mass in terms of the observed properties of the early peak of the light curve,
\beq\label{dm_peak}
 \Delta m_{\rm ej,peak} \approx 5.6 \times 10^{-3} M_\odot \left(  \frac{\tau_{\rm peak}}{8 {\rm \ d}} \right)^{2} \left(  \frac{v_{\rm ej}}{370 \ {\rm km \ s}^{-1}}\right).
\eeq
This ejected mass is large compared to the minimum obscuring mass estimated above (as is evident from the fact that the light curve's rise to initial peak is much longer than the light crossing time). The total kinetic energy carried by the outflowing shell of material, $E_{\rm K} = {1\over 2}  m v^2$, is
\beq\label{ek}
E_{\rm K,peak} \approx 7.6 \times 10^{45} {\rm  \ erg} \left(  \frac{\tau_{\rm peak}}{8 {\rm \ d}} \right)^{2} \left(  \frac{v_{\rm ej}}{370 \ {\rm km \ s}^{-1}} \right)^3 .
\eeq
We note that because of the velocity-cubed behavior of this expression, the kinetic energy estimate is relatively sensitive to the ejecta velocity uncertainty. To illustrate this more quantitatively, the $\pm 240$~km~s$^{-1}$ error bars of equation \eqref{williams_v} allow for an ejecta velocity of up to 610~km~s$^{-1}$ within one sigma, which gives $E_{\rm K,peak} \approx 3.4 \times 10^{46} {\rm  \ erg}$ in equation \eqref{ek}.

The kinetic energy estimate of equation \eqref{ek} can be compared to the radiated energy in the peak portion of the light curve, as plotted with a dot-dash line in the lower panel of Figure \ref{fig:erad}. 
The material must expand before it becomes transparent. In the absence of additional heating or cooling, adiabatic expansion decreases the internal energy -- and thus also the radiated energy -- relative to the kinetic by $(R_{\rm bb}/r_0)^{-1}$ for radiation pressure dominated ejecta. A scaling of $(R_{\rm bb}/r_0)^{-2}$ is appropriate for gas pressure dominated material.  However, if a even small fraction of the material recombines, this can lead to a much shallower decay in temperature or internal energy with radius \citep[see, for example, Figure 2 of][]{2010ApJ...714..155K}. 
For any of these possibilities the radiated energy of the ejecta should be lower than the kinetic energy. 

For a reddening of $\ebv=0.12$~mag, the radiated energy during the early-peak portion of the lightcurve is $\approx 2.0\times10^{45}$~erg.  
The kinetic energy estimated in equation \eqref{ek} is thus a factor of $\sim 4$ larger than the radiated energy. 
If the velocity of the ejecta is  higher than the nominal value, then this ratio increases. For $v_{\rm ej}=610$~km~s$^{-1}$, the kinetic energy exceeds the radiated by a factor of $\sim 17$. 
Using constraints on $r_0$ that will be derived in Section \ref{sec:progenitor}, ratios of $R_{\rm bb}/r_0 \sim 10$ are typical. This comparison shows that the implied ratios of kinetic to radiated energy are thus of the right order of magnitude, with a preference for $v_{\rm ej}$ somewhat higher than 370~km~s$^{-1}$, though still within the 1$\sigma$ limits of equation \eqref{williams_v}.

\subsubsection{Plateau}\label{sec:plateau_scalings}
The second phase of the light curve morphology, the plateau, exhibits relatively constant bolometric luminosity for $\sim40$~d and spans from 10 to 50~d after peak (as seen in the lower-left panel of Figure \ref{fig:phot}). 
The color evolution slows during this phase, and the effective temperature stabilizes in the $3-4\times 10^3$~K range (depending on reddening). 
These properties suggest that hydrogen recombination might be stabilizing the effective temperature as we observe a recombination wave propagating through the ejecta, lowering the opacity of material that recombines, as described by \citet{1993ApJ...414..712P} for type IIP supernovae. This idea was introduced in the context of LRN by \citet{2013Sci...339..433I} where it was invoked as a potential explanation of the light curve evolution in various LRN outbursts.

During the peak phase of the light curve, however, recombination energy is not an important contribution to the energetics. The photosphere is too hot, and the total energy, 
\beq\label{recomb}
E_{\rm recomb} \approx 2.9 \times 10^{44} {\rm erg} \left(  \frac{\Delta m}{0.01 M_\odot} \right),
\eeq
is too low by 1-2 orders of magnitude to explain the radiated energy given our estimates of the ejecta mass. The expression above describes the recombination potential energy per unit mass $\Delta m$ in a fully ionized plasma with solar composition abundances of H and He \citep{2013Sci...339..433I,2015MNRAS.447.2181I}.
In short, a recombination powered transient cannot simultaneously be of such short duration and as energetic as the early peak of M31LRN 2015 given the characteristic velocities observed. Thus, it appears that simple heat diffusion, rather than recombination power, serves as the primary regulator of the early light curve's temporal evolution. 

Returning to the plateau phase of the light curve, recombination energy could well play an important role in increasing the radiated energy (and modulating the photosphere location), in direct analogy to the evolution of type IIP supernovae -- though without nuclear decay heating \citep{1993ApJ...414..712P,2013Sci...339..433I} \citet{2013ApJ...769..109L} describe similar scaled-down type IIP transients from mass-loss in failed supernovae in which the energy is similar but the ejecta mass is much larger than M31 LRN 2015 -- giving lower ejecta velocities and different light curve evolution. 

The disappearance of H$\alpha$ emission lines from M31 LRN 2015's spectra taken during the plateau phase along with the drop in photosphere temperature indicate that material which was once recombining outside of the ejecta photosphere is now recombining at the photosphere. Hydrogen recombination is accompanied by a drop in ejecta opacity by several orders of magnitude. Because of this transition, material is often assumed to be completely transparent outside the recombination front and completely opaque inside of it. This implies that the photosphere is located at the same position as the recombination front, and that the effective temperature of the ejecta should stabilize to approximately the recombination temperature \citep[see, for example, the discussion in Section 2 of][]{2009ApJ...703.2205K}.

Interpreting the plateau phase of the lightcurve as a recombination-governed transient yields some insight into the properties of the ejecta. \citet{2015ApJ...805L..18W}  note that the lightcurve appears consistent with the ejection of $\sim 0.1 M_\odot$ by comparison to the luminosity and duration scalings of \citet{2013Sci...339..433I}. Appendix \ref{sec:recomb} elaborates on these same scalings for recombination transients. Here, we can use the observed properties of the recombination plateau to estimate the ejecta mass using equation \eqref{dmp_obs} and the ejecta's initial radius using equation \eqref{rip_obs}. Inserting $L_{\rm p} \sim 10^{39}$~erg~s$^{-1}$ for the plateau luminosity, $t_{\rm p}\sim 50$~d for the total duration, $T_{\rm rec} \sim 4000$~K for the recombination temperature (similar to the photosphere effective temperature), and $v_{\rm ej} \sim 370$~km~s$^{-1}$ for the ejecta velocity, we obtain 
\beq\label{dm_plateau}
\Delta m_{\rm ej,plateau} \sim 0.27 M_\odot,
\eeq
which scales as $\propto T_{\rm rec}^4 t_{\rm p}^4 v_{\rm ej}^3 \kappa^{-1} L_{\rm p}^{-1}$, and
\beq\label{Ri_plateau}
R_{\rm init} \sim 33 R_\odot,
\eeq
which scales as $\propto T_{\rm rec}^{-4} t_{\rm p}^{-2} v_{\rm ej}^4 \kappa L_{\rm p}^{2}$. For full details of these scalings, see Appendix \ref{sec:recomb} and \citep{2013Sci...339..433I}. Because of the high powers in these scalings, we caution that these estimates are uncertain at the factor of  a few level.

Even taken as order of magnitude estimates, a comparison of equations \eqref{dm_peak} and \eqref{dm_plateau} suggests that a small amount of mass ejected at somewhat higher velocity forms the early light curve peak, while later a larger amount of ejecta contributes to the plateau phase.

\section{Pre-Outburst Source}\label{sec:progenitor}

The detection of a progenitor source with known distance in the years prior to the transient outburst makes the lessons of M31LRN 2015 particularly useful. \citet{2015ApJ...805L..18W} report a source at the outburst location in {\it Hubble Space Telescope} (HST) imaging from August 2004, $\sim 10.5$~yr before the outburst took place.  F555W and F814W magnitudes were converted into $M_{V}$ and $V-I$ colors assuming a reddening of $\ebv=0.12\pm0.06$~mag to be $M_{V}=-1.50\pm0.23$~mag, and $(V-I)=1.05\pm0.15$~mag \citep{2015ApJ...805L..18W}. We transform these observations to a broader range of possible reddening values in what follows.  \citet{ 2015ATel.7173....1D} study a brightening of the progenitor using data from several ground-based surveys in the year prior to the outburst, and estimate a progenitor mass of $\sim 2-4M_\odot$. 

 In this section, we compare these detections  of the progenitor and precursor emission to detailed stellar models to attempt to infer the properties of the source over the decade prior to outburst. 
We will assume that the progenitor source is dominated by the a giant-star primary which has not yet been disturbed by its companion. To do so, we compare the source detection to post-main-sequence stellar tracks of non-rotating stars. 
These are simplifying assumptions -- we know that stellar evolution proceeds differently for stars in binary systems than for isolated stars \citep[e.g.][]{2016arXiv161103542D}. In particular, rapid rotation or phases of mass transfer can modify the colors of stars compared to their non-rotating single-star counterparts of the same mass \citep{2016arXiv161103542D}, and orbital instability can occur in some cases following such periods of mass transfer \citep{2002ApJ...565.1107P}. 
While a full consideration of these binary-star effects is beyond the scope of the modeling presented here, it may be worthwhile to consider binary-evolution scenarios for M31 LRN 2015 and other stellar merger sources in future work.

\subsection{Stellar Evolution Models}
To map the pre-outburst HST color and magnitude to the properties of a progenitor star we compute a grid of stellar models with the MESA stellar evolution code\footnote{version 7624} \citep{2011ApJS..192....3P,2013ApJS..208....4P,2015ApJS..220...15P}. 
We compute stellar tracks for stars of solar metallicity with masses between 2 and 8 $M_\odot$ from the pre main sequence until core helium ignition.

We use a modified version of the input lists provided to recreate Figure 16 of \citet{2013ApJS..208....4P} which are available on \texttt{mesastar.org}. The initial models have a composition of $Y = 0.272$ and $Z = 0.02$ and convection is determined by the Schwarzschild criterion using $\alpha_{\rm MLT} = 2$ and allows for 
exponential overshoot by 1\% of the scale height (\verb|overshoot_f|$=0.014$ and \verb|overshoot_f0|$=0.004$). 
 We include mass loss along the RGB through a \citet{1975MSRSL...8..369R} wind prescription with $\eta_{\rm R} = 0.5$ and do not include the effects of rotation. 
We calculate nuclear reactions using the JINA reaclib rates \citep{2010ApJS..189..240C}. We use standard OPAL opacities \citep{1996ApJ...464..943I} and we calculate surface properties using a grey atmosphere approximation.  

To map the MESA variables to a color and magnitude we use the MESA \texttt{colors} package, which follows the method outlined in \citet{1997A&AS..125..229L,1998A&AS..130...65L} to yield $UBVRI$ colors given surface temperature, surface gravity, and metallicity.

\subsection{Source Properties}

\begin{figure}[tbp]
\begin{center}
\includegraphics[width=0.48\textwidth]{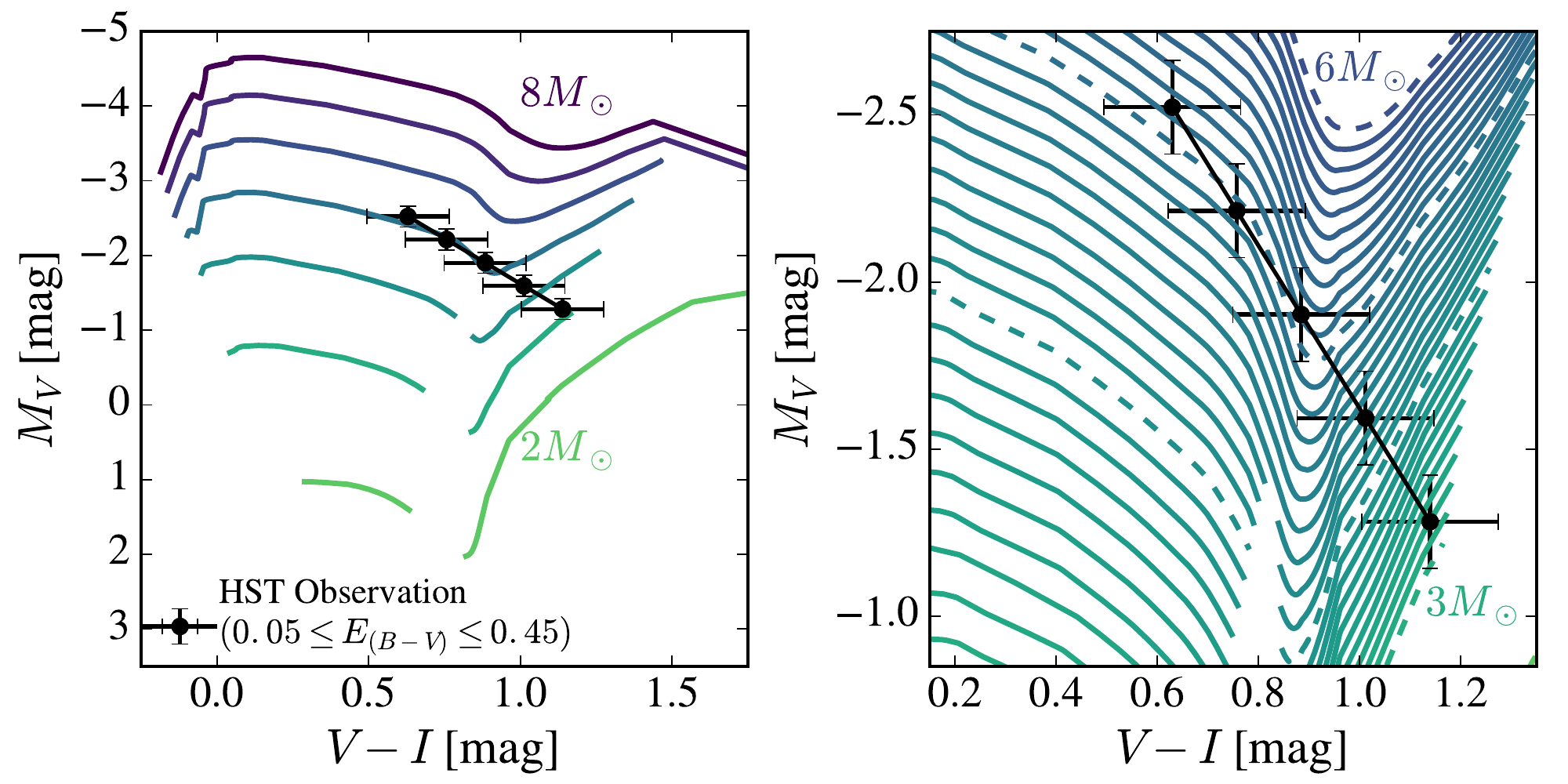}
\caption{Color-magnitude diagram comparing the HST-detected source to post-main sequence stellar evolutionary tracks computed in MESA. The tracks are masked to exclude portions of the evolution where the radius decreases (because an encounter could only occur as the primary star grows). The pre-outburst color and magnitude are re-computed from the \citet{2015ApJ...805L..18W} data to take $\ebv=$0.05, 0.15, 0.25, 0.35, 0.45~mag. Error bars include measurement and distance error, but not reddening error.  The left panel shows tracks of stars between 2 and $8 M_\odot$, in intervals of $1M_\odot$. The right panel zooms in and shows tracks with intervals of $0.1 M_\odot$ between 3 and $6M_\odot$ (integer values are plotted with dashed lines).  From these color magnitude diagrams, the pre-outburst source is clearly identified as subgiant star of several solar masses. }
\label{fig:cmd_tracks}
\end{center}
\end{figure}

In Figure \ref{fig:cmd_tracks}, we compare the HST detection to single-star evolutionary tracks in the $M_V$ versus $V-I$ color magnitude diagram (CMD). The tracks plotted are the post main sequence evolution of stars between $2M_\odot$ and $8M_\odot$.  We make the initial assumption that we can approximate the net emission as being solely due to the primary star.  The tracks are masked to only show portions of the evolution during which the stellar radius has not previously been larger. This selection is important in our present binary-merger context because a binary interaction can only occur while the evolving primary star is {\it growing} to larger radial extent.  

We transform the HST detection to a range of $\ebv$ values between 0.05~mag and 0.45~mag to span the range of uncertainty reported by  \citet{2015ApJ...805L..18W} and \citet{2015A&A...578L..10K}.  We plot error bars that include the distance modulus error and measurement error only (reddening error is not included). The reddening uncertainty thus implies a diagonal swath of possible color-magnitude pairs (not an uncorrelated error space).  

Figure \ref{fig:cmd_tracks} shows that the allowed region of CMD space selects portions of stellar tracks which range from $3M_\odot$ to about $5.5M_\odot$. The selected regions along these tracks always correspond with phases in which the primary star is growing in radius as it evolves.  This presents a picture very consistent with a star evolving to engulf its companion.  Further, it suggests that, unless the companion is in the very unlikely configuration of also being a giant, the light will be dominated by the much-more luminous evolved primary, which justifies our comparison to single-star tracks.

\begin{figure}[tbp]
\begin{center}
\includegraphics[width=0.45\textwidth]{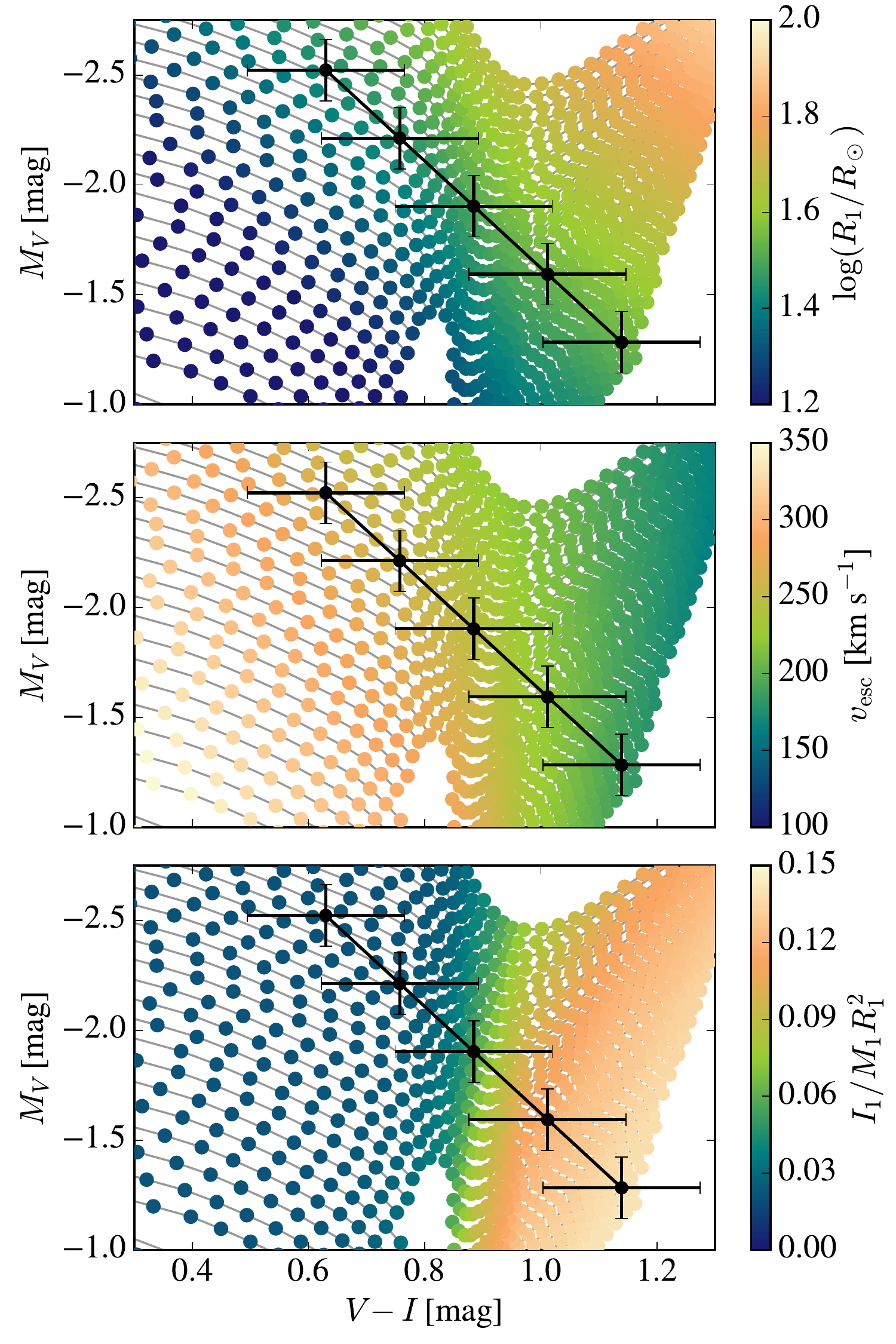}
\caption{Properties of pre-outburst progenitor models plotted along stellar evolutionary tracks. The source data are again plotted for $\ebv=$0.05, 0.15, 0.25, 0.35, 0.45~mag. Shortly before outburst, the primary star in the binary system was ascending the subgiant branch, during which its radius grows (top panel) and its escape velocity decreases (center panel). The specific stellar moment of inertia increases (lower panel) as the envelope's internal structure transitions from radiative to convective.   }
\label{fig:RIV}
\end{center}
\end{figure}

In Figure \ref{fig:RIV} we map the CMD space to the physical properties of the progenitor primary star, in particular radius, specific moment of inertia, and escape velocity.  
The panels of Figure \ref{fig:RIV} use colored dots along evolutionary tracks to show how the evolution of stars in the 3$M_\odot$ to 6$M_\odot$ range compare to the vector of progenitor source colors and magnitudes allowed by the HST data.  
The growth in radius along the post-main-sequence stellar tracks can be observed in the top panel of this diagram. Interestingly, the CMD swath allowed by the data selects objects within a factor of two in radius regardless of reddening. 
The stellar escape velocity $v_{\rm esc}$ is shown in the center panel of Figure \ref{fig:RIV}. The escape velocity decreases along the evolutionary tracks as stars go from their compact main sequence radii to extended giant-branch radii. 
 
The stellar moment of inertia is important in the context of binary star evolution because it relates to how much energy and momentum are needed to lock the primary star into corotation with its companion. We compute the moment of inertia from the radial profile of the stars' interior density profiles, $\rho_1(r)$, as
\beq\label{I1}
I_1 = {8 \over 3} \pi \int_0^{R_1} \rho_1(r) \ r^4 dr, 
\eeq
and define a specific moment of inertia $\eta_1 = I_1 / M_1 R_1^2$ \citep[e.g.][]{2006MNRAS.373..733S}.  This specific moment of inertia is plotted in the lower panel of Figure \ref{fig:RIV}. It changes by a factor of approximately seven in the CMD space selected by our HST measurements. In this transition we are seeing the sub-giant branch transition from models with primarily radiative envelopes (and low specific moments of inertia) to those with convective envelopes (with more mass at larger radii and thus higher specific moments of inertia).

\begin{figure}[tbp]
\begin{center}
\includegraphics[width=0.49\textwidth]{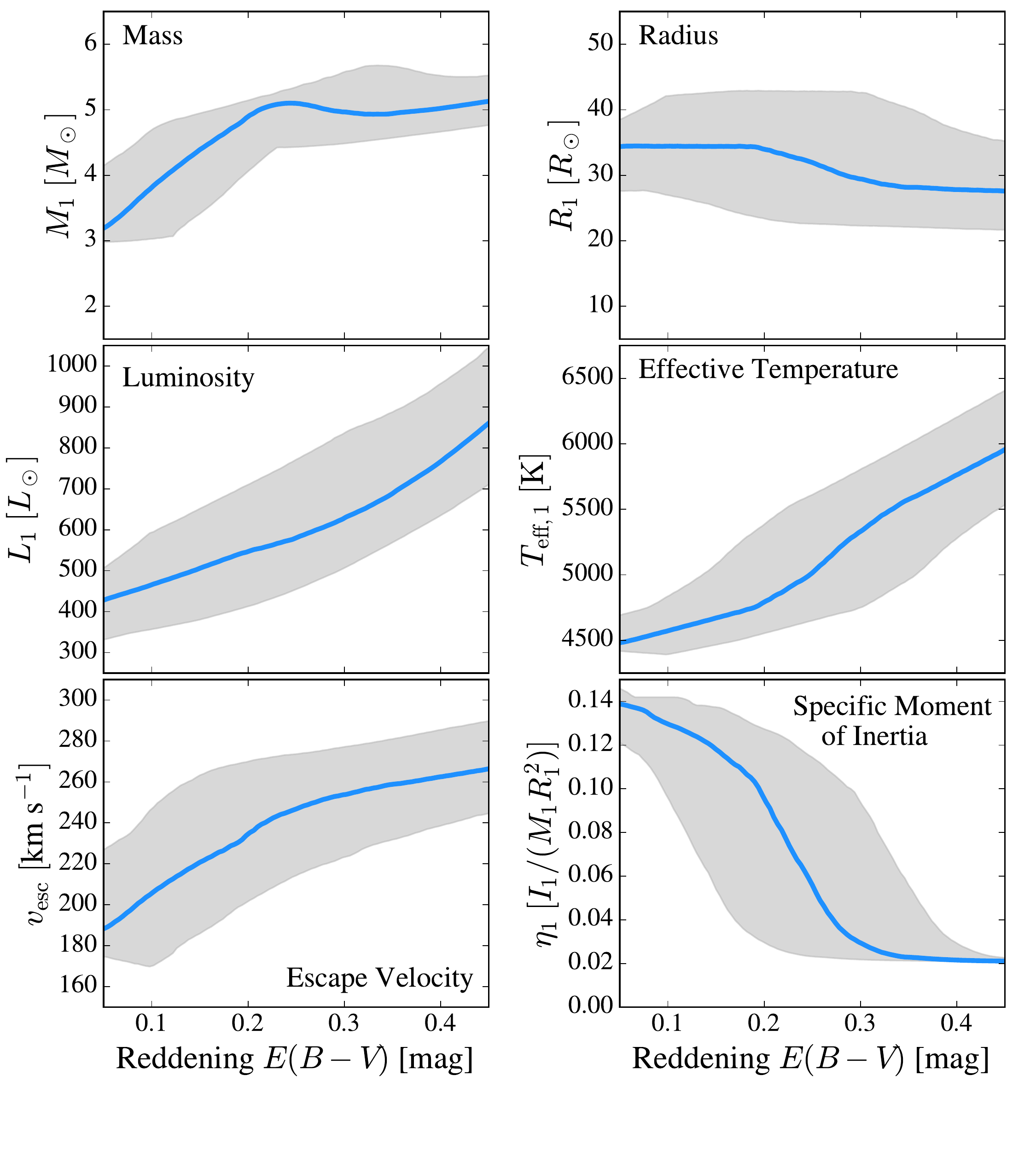}
\caption{Properties of pre-outburst progenitor models as a function of assumed reddening. These properties are measured from MESA models in the $M_V$, $V-I$ color-magnitude space allowed for a given reddening. Shaded regions show the range of properties within the $\pm 1\sigma$ photometric measurement error bars of Figures \ref{fig:cmd_tracks} and \ref{fig:RIV}.  The pre-outburst source is a 3 to 5.5$M_\odot$ star of approximately $35R_\odot$ and $\sim400-800 L_\odot$. It has a an escape velocity in the range of $\sim180-280$ km s$^{-1}$.  The object's internal structure and thus specific moment of inertia spans a wide range depending on reddening, at low reddening the star has a convective envelope and $\eta_1 \approx 0.14$, while at high reddening the star has a radiative envelope and lower specific moment of inertia of $\eta_1 \approx 0.02$. }
\label{fig:prog_prop}
\end{center}
\end{figure}

In Figure \ref{fig:prog_prop}, we derive numerical values for a few key properties of the progenitor primary star as a function of reddening. We plot derived mass, radius, specific moment of inertia, and escape velocity along with their $\pm1\sigma$ error regions. As in the previous figures, we assume fixed reddening values and consider the contribution of other sources to the error budget (no reddening error).  We find that the pre-outburst star is a $\sim 3-5.5 M_\odot$ star, with radius of $\sim25-40 R_\odot$.  These masses and radii imply typical escape velocities from the primary of $\sim180-280$ km s$^{-1}$.  As a function of the assumed reddening, the interior structure of the fitted progenitor star changes: if the reddening is $\ebv \lesssim 0.25$~mag, then the primary star has a convective envelope, and a relatively high specific moment of inertia, $\eta_1$. If, however, the reddening is higher, the primary envelope is radiative and has lower specific moment of inertia.

\subsection{Precursor Emission}

\citet{2015ATel.7173....1D} studied the progenitor using archival images from a number of sources including the CFHT, Local Group Survey, and SDSS. Their analysis shows a source consistent with being constant between 2002 and 2009, followed by a $~2.4$~mag brightening (compared to the HST imaging) to $g=20.8$~mag in October 2014. The date of the observed brightening is just 15 months prior to the outburst. This {\em precursor} emission marks an important signature of the oncoming transient. 

The brightening of 2.4 magnitudes implies a factor of $\sim 10$ brightening (in this wavelength range) of the progenitor. In general, this could be caused by a change of color or of overall luminosity (or, of course, a combination of the two). If the source has constant effective temperature, a factor of 10 increase in luminosity of the progenitor, $L_1 = 4\pi R_1^2 \sigma_{\rm b} T_{\rm eff,1}^4$, implies a factor $10^{1/2} \sim 3$ increase in photosphere radius $R_1$. If instead we imagine a source of constant photosphere radius, the blackbody effective temperature must increase to $\sim7500$K from  $4600$K (taking approximate parameters for $\ebv=0.15$) in order to explain the $g$-band flux increase. This source would have undergone a factor of $\sim7$ increase in bolometric luminosity. With the single color observation available it is difficult to distinguish between these possibilities (or a combination of radius and temperature increase).

It is, however, worthwhile to compare the implied radius and temperature increase to the scales at the first detection of the transient (for which multicolor photometry is available). The transient is first detected at a radius of $1$~AU, approximately a factor of $\sim6$ larger than $R_1$. The initial color of the transient is $\sim 5000$K. These scales show that the increase in luminosity could be accommodated by either increasing radius (at constant $T_{\rm eff}$) or by increasing temperature -- though not by increasing photosphere radius with a significant accompanying decrease in photosphere radius (which would appear to be inconsistent the growing photosphere seen during the transient phase).   

\section{A Model For System and Transient Together}\label{sec:comb_req}

\subsection{Combined Properties}
M31LRN 2015 was identified as a stellar merger by \citet{2015ApJ...805L..18W} and \citet{2015A&A...578L..10K}  through comparison of the outburst properties to similar optical transients like V838 Mon and V1309 Sco, which have been associated with a stellar merger origin \citep{2006A&A...451..223T,2006MNRAS.373..733S,2011A&A...528A.114T,2014ApJ...788...22P,2014ApJ...786...39N}.  
The pre-outburst progenitor source appears consistent with this hypothesis. The HST detection places the sub-giant primary star in a portion of the CMD in which it is evolving and growing in radius -- perhaps to engulf a companion.

A comparison of the progenitor source and the outburst itself allows us to draw several conclusions about this process.  The optical transient rises from discovery to peak in a timescale of $\sim 8$~d.  Depending on the reddening, the orbital period of a test mass at the surface of the primary varies from $\sim 13$ to $7$~d (for low to high $\ebv$, the primary's dynamical timescale, $\left(R_1^3 / G M_1 \right)^{1/2}$, is $2$ to $1.2$~d, respectively). Thus the initial transient outburst is quite rapid relative to the binary orbital period (of the same order) and the entire optical transient transpires over only tens of orbits of the binary. Precursor emission from October 2014 marks the first detection of a brightening 15 months, or $\sim30-70$ orbits prior to merger.

Although the ejecta giving rise to the optical transient appear to be liberated on a dynamical timescale in the merger, the mass ejected is only a small fraction of the total system mass. In  Section \ref{sec:outburst} we estimated that of order $10^{-2} M_\odot$ is ejected in the early peak of the outburst light curve (the portion with timescale similar to the orbital period). Of order $0.3 M_\odot$ may be ejected in the plateau portion.  These ejecta masses are small by comparison to our estimated primary-star masses from the progenitor imaging. 

\begin{figure}[tbp]
\begin{center}
\includegraphics[width=0.45\textwidth]{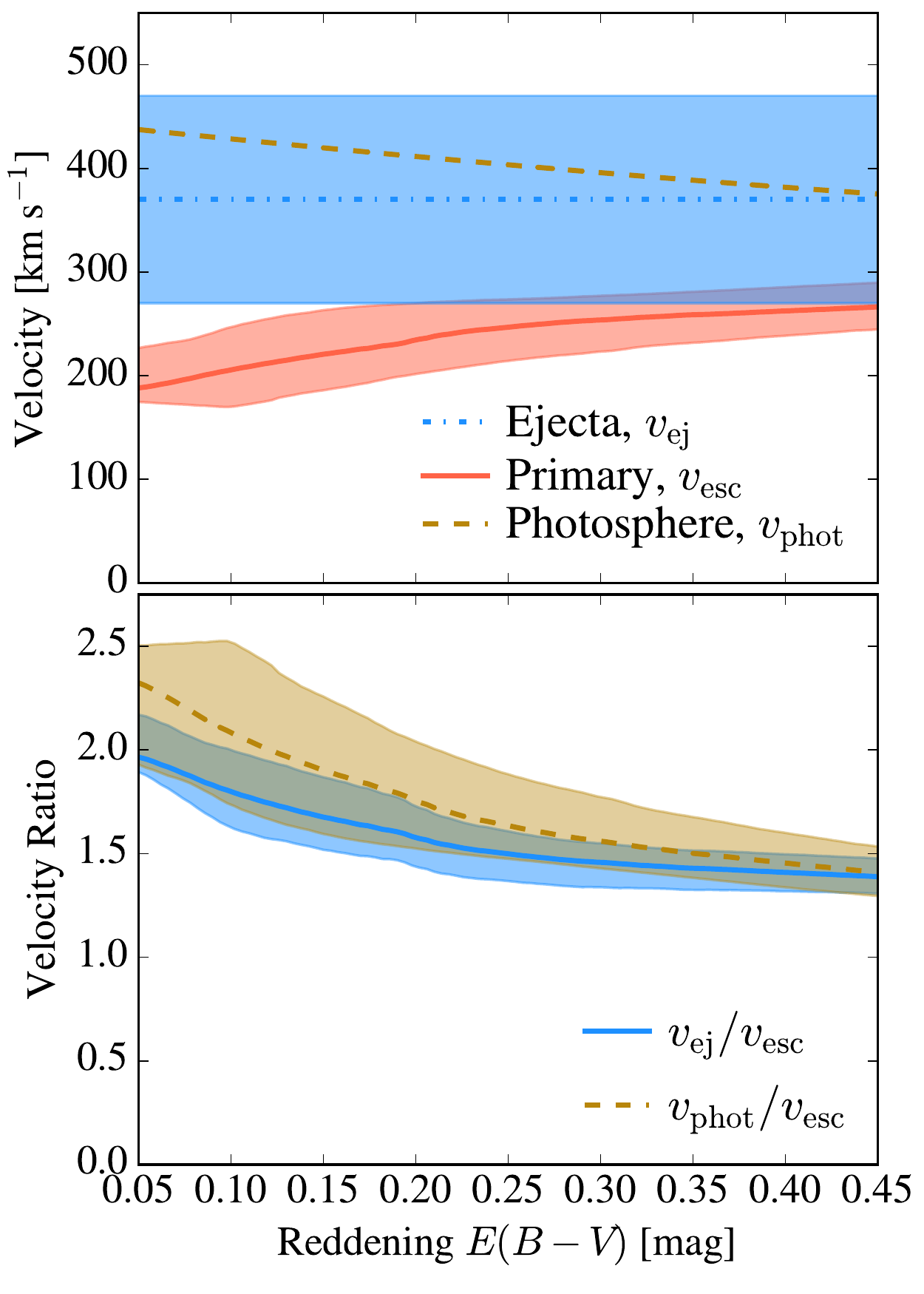}
\caption{Taking the outburst and pre-outburst system together, this figure compares characteristic velocities. In the top panel, we plot the ejecta velocity inferred from spectral emission-line FWHM (blue), to the escape velocity of the primary (red), and the expansion velocity of the photosphere.  In the bottom panel,  the blue, solid line shows the ratio of the ejecta velocity to the escape velocity, while the dashed gold line shows the ratio of the  photosphere expansion velocity to the escape velocity. These comparisons show that the ejecta expand at velocities similar to or slightly larger than  the escape velocity $(v_{\rm ej}\gtrsim v_{\rm esc})$.   }
\label{fig:vel}
\end{center}
\end{figure}

 A final useful point of comparison lies in the characteristic velocities of the ejecta as compared to the progenitor system escape velocity. We have two observational handles on the velocity of the ejecta. The first is the H$\alpha$  line FWHM from the \citet{2015ApJ...805L..18W} data near the peak of the optical transient, equation \eqref{williams_v}. A second velocity measure is the expansion velocity of the photosphere as computed in Figure \ref{fig:phot}. 
 In Figure \ref{fig:vel} we plot these characteristic velocities describing the progenitor system and outburst along with their ratio to the primary's escape velocity. Across the range of possible source reddening values, the ejecta outflow at velocities similar to or greater than the primary star's escape velocity $(v_{\rm ej} \gtrsim v_{\rm esc})$. 
From this analysis we therefore can glean that in the M31LRN transient, we are observing the rapid, dynamical ejection of a small portion of the system mass. The characteristic masses are small relative to the system mass, the velocities are of the same order as the system escape velocity, and the timescale is similar to the primary's dynamical time.  
 
\subsection{Proposed Interpretation}
In this section, we propose an association of the modeled properties described above with various phases of the common envelope interaction. 

\subsubsection{Precursor: Pre-Common Envelope Interaction}

The precursor brightening is likely to be associated with the oncoming merger event (which in turn, generates the luminous optical transient). The detection arises just tens of orbits prior to merger, so it is appealing to imagine the interaction ramping up toward runaway during this phase. The cause of the precursor emission seems likely to be some early tidal heating and disturbance within the primary-star atmosphere through interaction between the secondary star and the outermost layers of the primary. Another possibility, recently outlined by \citet{2016MNRAS.461.2527P}, is that cold, bound material originally lost from the outer, $L_2$, Lagrange point in binary systems could cause a brightening (and increase in temperature) as it falls back and virializes. These possibilities are conceptually similar -- involving some heating of the surface layers due to the presence of the companion -- and both appear to be possible given the limited evidence from the precursor emission.

It is useful to compare the luminosity during this phase of $\sim7-10L_1 \sim 10^{37}$~erg~s$^{-1}$ to the orbital energy of the presumed binary with separation $a\sim R_1$. This energy scale is $E_{\rm orb} = G M_1 M_2 / a \sim G M_1 M_2 / R_1 \sim 5\times10^{47} \left(M_2/M_\odot\right)$~erg. If the radiated luminosity during the precursor phase represents orbital decay, and has a radiative efficiency $\eta$, then the timescale for inspiral is $\tau_{\rm decay} \sim E_{\rm orb} / \left( L/\eta \right)$. For $\eta=0.1$ and the energy and luminosity above, the implied orbital decay timescale is $\tau_{\rm decay} \sim 30$~yr, or $\sim 10^3$ orbital periods.  The precursor emission alone does not account for the onset of merger within approximately a one year timescale, and a subsequent runaway to higher energy loss rates is needed.

\subsubsection{Early Peak: Shocked Ejecta from the Onset of Common Envelope}\label{interpret:peak}
The M31LRN 2015 light curve shows an early peak, analyzed in Section \ref{sec:early_peak_scalings}. We argue that a small amount of mass, of order $10^{-2} M_\odot$, ejected at velocities similar to or exceeding the escape velocity, best explains this signature. Due to the small mass and rapid timescale, we associate these ejecta with the first phase of contact of the merging binary, where the relative velocity of the secondary across the primary's atmosphere drives shocks through the primary's outer envelope \citep[e.g.][]{2006MNRAS.373..733S,2012MNRAS.425.2778M}. 
In this scenario shocks are generated by the relative motion of the secondary through the primary's envelope.

Material from the stellar envelope would be gravitationally focused toward the secondary as it skims through the stellar atmosphere.  This gravitational focussing effect from the strong density gradient of the stellar limb creates a flow morphology in which material is shock-heated and a fraction is ejected radially outward \citep[see, for example, the flow morphology in simulations presented by][]{2015ApJ...803...41M,2015ApJ...798L..19M}.   These shocks heat and accelerate a small portion of the mass to high velocities as they run down the density gradient of the primary's disturbed envelope.  The observed presence of broad H$\alpha$ emission during this phase of M31 LRN 2015 indicates that hot, ionized gas is recombining outside the photosphere and is consistent with a scenario that generates hot, fast initial ejecta. The eventual ejecta in such a scenario will have a distribution of velocities of the same order, or faster than, the escape velocity from the primary star's envelope. 

As a check on this scenario, we can use the scalings of Section \ref{sec:early_peak_scalings} and the size of the primary, $R_1$, as the ejecta's origin to estimate a peak luminosity of 
\beq\label{Lp}
L_{\rm peak} \approx 1.0 \times 10^{39} {\rm erg \ s}^{-1} \left(  \frac{v_{\rm ej}}{370 \ {\rm km \ s}^{-1}} \right)^2  \left(  \frac{ R_1 }{35 R_\odot} \right), 
\eeq
that results from the ejection of a shell of mass of $\Delta m_{\rm ej,peak}$ with velocity $v_{\rm ej}$. We have assumed $L_{\rm peak} = E_{\rm int}(t_{\rm peak}) /t_{\rm peak}$, that the original internal energy is similar to the ejecta kinetic energy $(E_{\rm int}(0) = E_{\rm K,peak}$, as would be the case if the material were strongly shocked) and that the internal energy declines as the gas expands with $(R_{\rm bb}/r_0)^{-1}$ where $r_0=R_1$ \citep[e.g.][]{2001thas.book.....P, 2009ApJ...703.2205K}. 
Despite the many crude ingredients that form the basis of this calculation, this estimate agrees with the bolometric light curve of Figure \ref{fig:phot}. This agreement suggests that the ejection of a small amount amount of mass from the stellar surface can produce the early transient light curve.

\subsubsection{Later Outflow: Embedded Phase and Merger}

When the secondary star is deeply embedded within the envelope of the primary, gravitational interaction drives high density material from the stellar interior outward \citep{2015ApJ...803...41M}.  In contrast to the phase of early contact, at the onset of common envelope, this material is not free to expand into the low-density stellar atmosphere but instead thermalizes on its surroundings, effectively sharing the orbital-energy (drained from the secondary's orbit) with a large amount of surrounding primary-star envelope material \citep[see, for example, the later stages of the simulations of ][]{2014ApJ...786...39N}.  The continuous stirring of the envelope by the inspiralling-secondary is a violent process, but this phase drives a slower, albeit more massive outflow \citep[as, for example, seen in simulations by][]{1989ApJ...337..849T,1998ApJ...500..909S,2016MNRAS.458..832S,2016MNRAS.tmp.1479I}. Because the energy deposited by the secondary is shared with more mass in the embedded phase, the specific energy of the ejecta is lower and the imprint of these material on the light curve is longer, lower-temperature, and of lower peak luminosity. 

In this case, the photosphere is at larger radii (and lower temperatures) implying that the material is able to recombine prior to becoming transparent.  The H$\alpha$ emission lines in the spectra fade, and the photosphere temperature stabilizes as hydrogen recombination controls the opacity.  In this phase, the lightcurve behaves comparably a scaled-down type IIP supernova \citep{1993ApJ...414..712P,2013Sci...339..433I}.
Using the scalings of \citet{2013Sci...339..433I} explored in Appendix \ref{sec:recomb} we estimate an ejecta mass of order $\sim 0.3 M_\odot$. 
The initial radius of mass ejection estimated using this approach appears compatible with the primary star's radius as derived from the $HST$ progenitor detection, since both are of order $\sim 35 R_\odot$.

Finally, the photosphere begins to recede after day 40 in the lower-right panel of Figure \ref{fig:phot} suggesting that mass ejection has slowed or shut off at this time. We interpret this transition as the end of the binary merger: following a phase of orbital inspiral, the secondary tidally disrupts inside the primary envelope, leaving a merged remnant.  Ongoing observations of the later phases of the transient (and its subsequent evolution in later stages) will allow us to constrain the total heat deposited into the primary's envelope and to follow its resultant relaxation as it cools.

\section{The Onset of Common Envelope: Pathways and Observational Implications}\label{sec:mergers}

Multiple  mechanisms of angular momentum loss can destabilize a binary system and drive it toward merger. Two of these are Roche lobe overflow through the outer Lagrange point, which carries mass and angular momentum away from the system,  and the Darwin tidal instability, which deposits orbital angular momentum into the reservoir of the primary's envelope \citep{1879RSPS...29..168D}.  
Either angular momentum loss channel destabilizes and desynchronizes the orbital motion of the secondary from the rotation of the primary envelope leading the two objects to plunge together with significant velocity shear at onset of common envelope (when the binary separation is equal to the radius of the primary, $a=R_{1}$). However, they seem likely to differ in the hydrodynamics of the initial mass ejection and, as a result, in the transients they should give rise to.

\begin{figure*}[tbp]
\begin{center}
\includegraphics[width=0.95\textwidth]{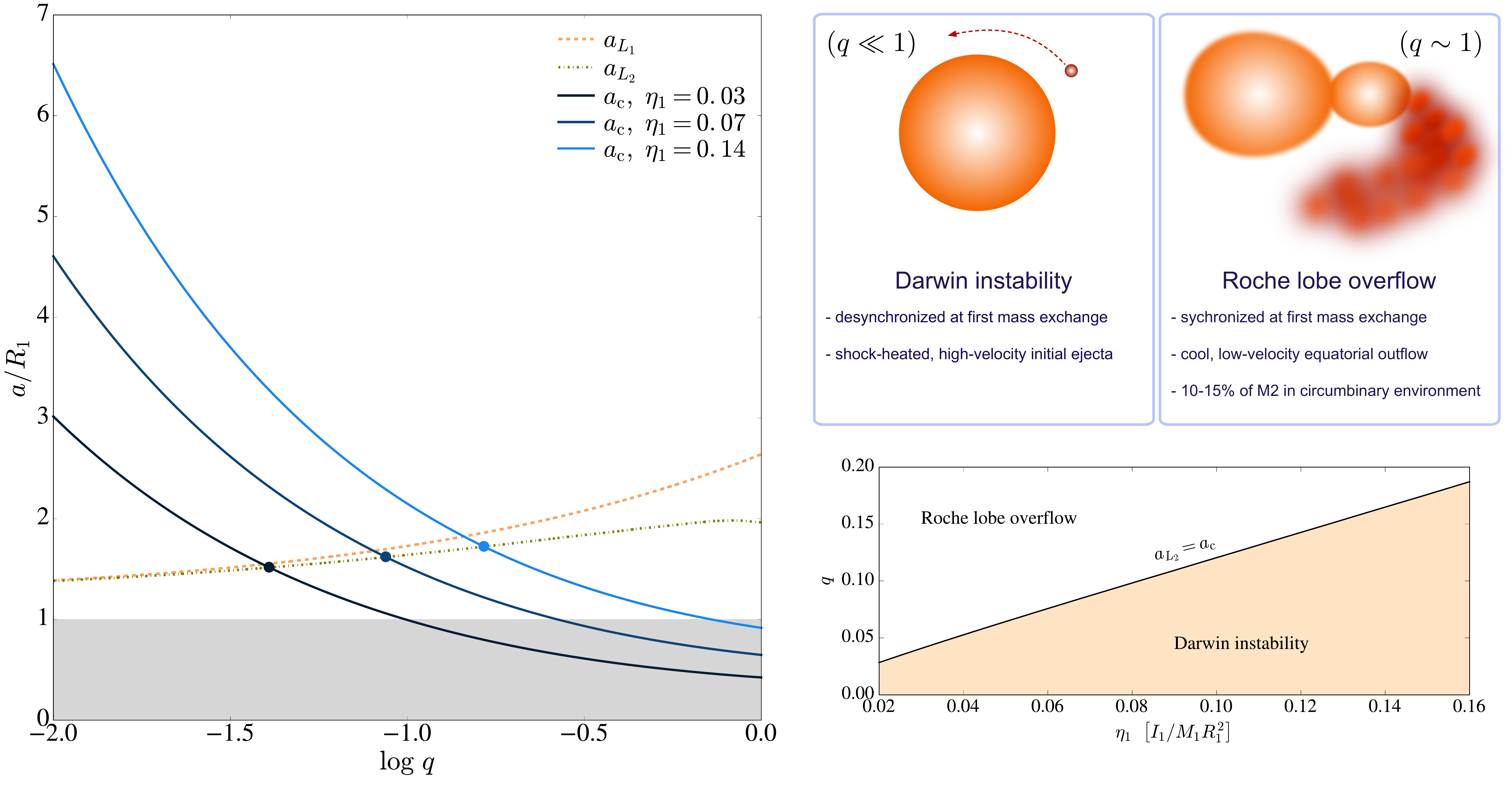}
\caption{Pathways to the onset of runaway interaction in a binary system. 
The left panel shows characteristic separations of orbital de-synchronization. Two possible processes, the Darwin tidal instability and non-conservative $L_2$ Roche lobe overflow can drive a binary pair toward merger. Points mark the transition where $a_{L_2}=a_{\rm c}(\eta_1)$.  The lower-right panel shows this transitional mass ratio for a range of primary-star specific moments of inertia, $\eta_1$. 
 Systems with $q\ll1$ are likely to merge via the Darwin instability, while systems with $q\sim1$ maintain corotation until mass transfer commences and merge by Roche lobe overflow. The pathway a merging system takes may have implications for the observable properties its transient, as outlined in the upper-right hand side panels.  
}
\label{fig:sep}
\end{center}
\end{figure*}

\subsection{Roche Lobe Overflow Leading to $L_2$ Mass Loss}

If the primary star is locked into corotation with the secondary, it can overflow its Roche Lobe at the $L_1$ (inner) Lagrange point  as it evolves toward contact. If the ensuing  mass transfer is, or becomes, unstable \citep[e.g.][]{1987ApJ...318..794H,1997A&A...327..620S,2002ApJ...565.1107P,2006ApJ...643..381D,2011ApJ...739L..48W,2012ApJ...760...90P,2015MNRAS.449.4415P}, the mass transfer rate can run away until the system also overflows the $L_2$ (outer) Lagrange point \citep{1979ApJ...229..223S,1998CoSka..28..101P,2016MNRAS.455.4351P,2016MNRAS.461.2527P}. Mass lost from the $L_2$ point carries angular momentum away from the bound pair, driving them toward merger. 
 
In Figure \ref{fig:sep}, we plot the separation, $a$, at which the primary undergoes Roche lobe overflow and begins to transfer mass onto the secondary (assuming corotation), this is approximated by \citet{1983ApJ...268..368E}'s formula,
\beq\label{RocheLimit}
{a_{\rm L_1} \over R_1} = \frac{0.6 q^{-2/3} + \ln(1+q^{-1/3}) }{0.49 q^{-2/3}},
\eeq
where $q=M_2/M_1$ is the binary mass ratio.
\citet{1983ApJ...268..368E}'s formula is an approximation of the separation at which the volume of the primary's Roche lobe is equal to the volume of the unperturbed star, $4/3 \pi R_1^3$. 
We perform a conceptually similar calculation to determine the approximate separation at which the system will overflow the $L_2$ Lagrange point, $a_{L_2}$. To compute this value, we compare the volume enclosed by the equipotential surface passing through $L_2$ (limiting our integration to the primary's lobe) to the primary's effective volume. We plot $a_{L_2}$ along with $a_{L_1}$ in the left panel of Figure \ref{fig:sep}. We find that $a_{L_2}$ is always smaller than $a_{L_1}$, with the difference  most substantial for $q\sim1$.  
In Appendix  \ref{sec:dml2}, we compute the mass loss needed to bring the binary from the point where it begins to shed material ($a=a_{L_2}$) to the onset of common envelope ($a=R_1$), and find that this mass is 10-15\% of $M_2$ across a range of mass ratios $q$, as shown in Figure \ref{fig:dml2}.

\subsection{Darwin Tidal Instability}
Binary systems with low-mass secondaries are subject to an orbital instability known as the Darwin instability \citep{1879RSPS...29..168D}. 
As the primary's envelope grows, a situation can arise where the angular momentum budget of the secondary's orbital motion is too small to lock the primary's envelope into corotation. As desynchronization ensues, angular momentum is drained from the secondary object's orbit and the orbit decays on a tidal-dissipation timescale.
The condition for this instability can be written in terms of the moments of inertia of the orbit $I_{\rm orb}$ and the moment of inertia of the primary, $I_1$, as \citep{2001ApJ...562.1012E}, 
\beq
I_{\rm orb} \lesssim 3 I_1.
\eeq
 In the above expression, $I_{\rm orb} = \mu a^2$, where $\mu = M_1 M_2 / M$ is the reduced mass, $M=M_1+M_2$ is the total mass, and $a$ is the orbital separation.  The primary's moment of inertia can be written as $I_1 = \eta_1 M_1 R_1^2$; it is computed from the interior structure of the primary using equation \eqref{I1}.  Though the moment of inertia of the secondary could enter into the above expression \citep{1980A&A....92..167H}, we ignore it here under the assumption of an evolved star primary with a more compact companion. 

The critical separation which leads to tidal instability can be written, 
\beq\label{ac}
a_{\rm c} =  R_1 \sqrt{3 \eta_1 \left( 1 + q^{-1} \right) }.
\eeq 
In the left-hand panel of Figure \ref{fig:sep}, we compare critical separations to the radius of the primary and the separations of Roche lobe overflow for synchronized systems, $a_{L_1}$ and $a_{L_2}$. We use three different specific moments of inertia, $\eta_1$, which span the range of typical stellar values shown in Figure \ref{fig:prog_prop} (low $\eta_1$ is relevant for main-sequence stars, while evolved stars with convective envelopes have higher $\eta_1$). 
The lower-right panel of Figure \ref{fig:sep} delineates the transition in mass ratio (as a function of primary-star specific moment of inertia, $\eta_1$) between Darwin unstable systems and Roche lobe overflow systems as divided by the line $a_{\rm c} = a_{L_2}$. 

Once systems desynchronize through the Darwin instability, they decay on a tidal friction timescale, which scales as $(a/R_1)^8$. As a result, we can expect that the orbital decay starts very slowly but becomes increasingly rapid as the binary enters the nonlinear regime of $a\sim R_1$. For a recent discussion of these decay timescales in the context of the V1309 Sco binary merger transient, see the discussion section of \citet{2014ApJ...786...39N}.

\subsection{Possible Consequences of Pathway for Merger Transients}\label{sec:obs_pathways}

A comparison of the desynchronization radii plotted in Figure \ref{fig:sep} shows that in some cases the system becomes Darwin unstable at larger separations than those for mass transfer, while in other cases unstable Roche lobe overflow proceeds to $L_2$ mass loss. We explore the differences between those pathways here and in the cartoon panels in Figure \ref{fig:sep}.

In systems that merge through unstable Roche lobe overflow, the initial ejecta are trailed off gently in a spiral wave from the $L_2$ point.  Beyond a few to ten times the orbital semi-major axis, these material pile up into a wind-like outflow with $\rho \propto r^{-2}$ \citep{2016MNRAS.455.4351P} in other cases of the binary mass ratio, the bulk of the material might remain bound to the binary \citep{2016MNRAS.461.2527P}.  In general, the material is cool, and the radial component of the ejected material's velocity is small, with the maximum velocity at infinity being about 25\% of the system escape velocity,$v_\infty \lesssim 0.25 v_{\rm esc}$ \citep{1979ApJ...229..223S,2016MNRAS.455.4351P}.  

 Mass loss from $L_2$ begins slowly but eventually exponentiates leading to the runaway merger. 
This transition takes place over many orbital periods, and, with a sufficient column depth, the photosphere could move outward into the ``wind''.  With photosphere radii larger than the overlap of the spiral wave of mass-loss (a few and ten times the binary separation), the luminosity would increase with the radiated energy originating in the relative velocity of spiral features \citep{2016MNRAS.455.4351P}. 
The characteristic features of transients generated by systems brought to merger through Roche lobe overflow will therefore be a mass-rich circumbinary environment (containing a total mass of 10-15\% $M_2$, preferentially concentrated in an equatorial wedge of uncertain $H/R$), the potential for precursor emission if the photosphere moves past the internal shock radius, and low ejecta velocities (relative to the system escape velocity). 

Systems that are driven to merger through the Darwin instability desynchronize when there is no longer sufficient angular momentum in the secondary's orbit to maintain the corotation of the primary. Thus, they proceed toward merger (on a tidal-dissipation timescale) without significant mass loss \citep{2001ApJ...562.1012E,2014ApJ...786...39N}. These systems may decay more rapidly in their late pre-merger phase according to evidence from SPH simulations \citep{2014ApJ...786...39N}.
When these systems reach separations similar to the mass-transfer separation for synchronized systems, $a_{L_1}$, there is a lack of corotation. The secondary would sweep through the atmosphere of the primary with high mach number \citep[e.g.][]{2015ApJ...803...41M}, shock heating and ejecting a fraction of the primary star's outermost layers,  primarily in the orbital plane \citep[e.g. Figure 2 of][]{1989ApJ...337..849T}.    This flow morphology is therefore quite different that the cold overflow from one potential well to another that is possible in a synchronized system.\footnote{This effect can also be phrased in terms of the effective equipotential surfaces experienced by fluid that is not corotating with the binary -- the potential is higher at the saddle points $L_1$ and $L_2$ because of the additional kinetic energy than in a corotating system \citep{2007ApJ...660.1624S}. }
We postulate that without the dense circumbinary outflow that characterizes Roche lobe overflow systems, this initial ejecta expands uninhibited, with characteristic velocities of order, or larger than, the system escape velocity.

\section{Was M31LRN 2015  A Darwin Unstable Merging Pair?}\label{sec:DI}

\subsection{Argument For a Darwin Unstable Merger}
A key characteristic of the M31 LRN 2015 transient is the high velocity of the initial ejecta as compared with the system escape velocity ($v_{\rm ej} \gtrsim v_{\rm esc}$, Figure \ref{fig:vel}). An appealing possible interpretation of this evidence is that the system lacks the dense, slowly-expanding circumbinary environment that $L_2$ mass loss deposits. For all secondary masses larger than $0.1M_\odot$, the total mass in such an outflow ($\gtrsim 0.1M_2$) would be large compared to the initial ejecta mass inferred in section \ref{sec:outburst}, implying that the slow-moving (and massive) early ejecta would decelerate any later, fast moving ejecta.  The apparent preservation of high velocities in the early peak ejecta of M31 LRN 2015 therefore implies that the binary may have come to the onset of common envelope through the Darwin tidal instability channel. 

With current evidence it is impossible to say full certainty whether such an interpretation is correct or unique. Another possible interpretation is that the early and late ejecta have different geometries, and our sightline is oriented such that we see the faster ejecta escaping. This is certainly possible, but to first order we expect the bulk of pre-merger and merger ejecta to be concentrated in the orbital plane, ensuring their interaction \citep[e.g.][]{1989ApJ...337..849T,2008ApJ...672L..41R,2012ApJ...744...52P, 2016MNRAS.455.4351P}, though \citet{2006MNRAS.365....2M} show that late ejecta might be collimated in the polar direction by these equatorial overdensities. 

Despite these uncertainties, if the system merged via the Darwin instability it implies several interesting constraints on the mass of the secondary object. We work through those constraints in this section in their application to M31 LRN 2015, but emphasize the applicability of this discussion to other systems should future evidence reveal their merger channel.

\subsection{Implied Constraints on $M_2$}
In this section, we consider what limits may be placed on the  mass of the unseen secondary star that was engulfed to create M31LRN 2015. 
We use the properties and energetics of the transient outburst along with our discussion above of the possible pathways to merger in order to consider the constraints on the properties of the merging binary.  

\subsubsection{An Upper Limit: Merger Pathway}

The requirement that the system merge via the Darwin instability has the effect of placing an upper limit on the mass ratio, $q$. 
Because the pre-outburst HST imaging constrains the primary mass and structure, it is also possible to place an upper limit on the mass of the secondary, $M_2$.  
We note that the pathway to merger is an upper limit on the mass ratio only for Darwin-unstable systems.  Were a transient's observational properties consistent with a Roche lobe overflow pathway to merger, this would instead imply a lower, rather than upper, limit on the binary mass ratio. 

To apply this constraint, we specify that the system must destabilize due to the Darwin instability at greater separation that it begins to lose mass from $L_2$, $a_{\rm c} \ge a_{L_2}$. This implies a maximum allowed $q$ for a given $\ebv$ reddening, which gives $a_{\rm c} = a_{L_2}$.  As can be seen in equation \eqref{ac} and Figure \ref{fig:sep}, this constraint depends on the specific moment of inertia of the primary star's envelope, $\eta_1$.  We therefore apply the constraints derived in Section \ref{sec:progenitor} and plotted in Figure \ref{fig:prog_prop}.  
For lower $q$, the pair becomes Darwin unstable at larger separations  $(a_{\rm c} > a_{L_2})$. However, for higher masses, the system remains synchronous until mass loss from $L_2$ begins.

\subsubsection{A Lower Limit: Energetics}

\begin{figure}[tbp]
\begin{center}
\includegraphics[width=0.45\textwidth]{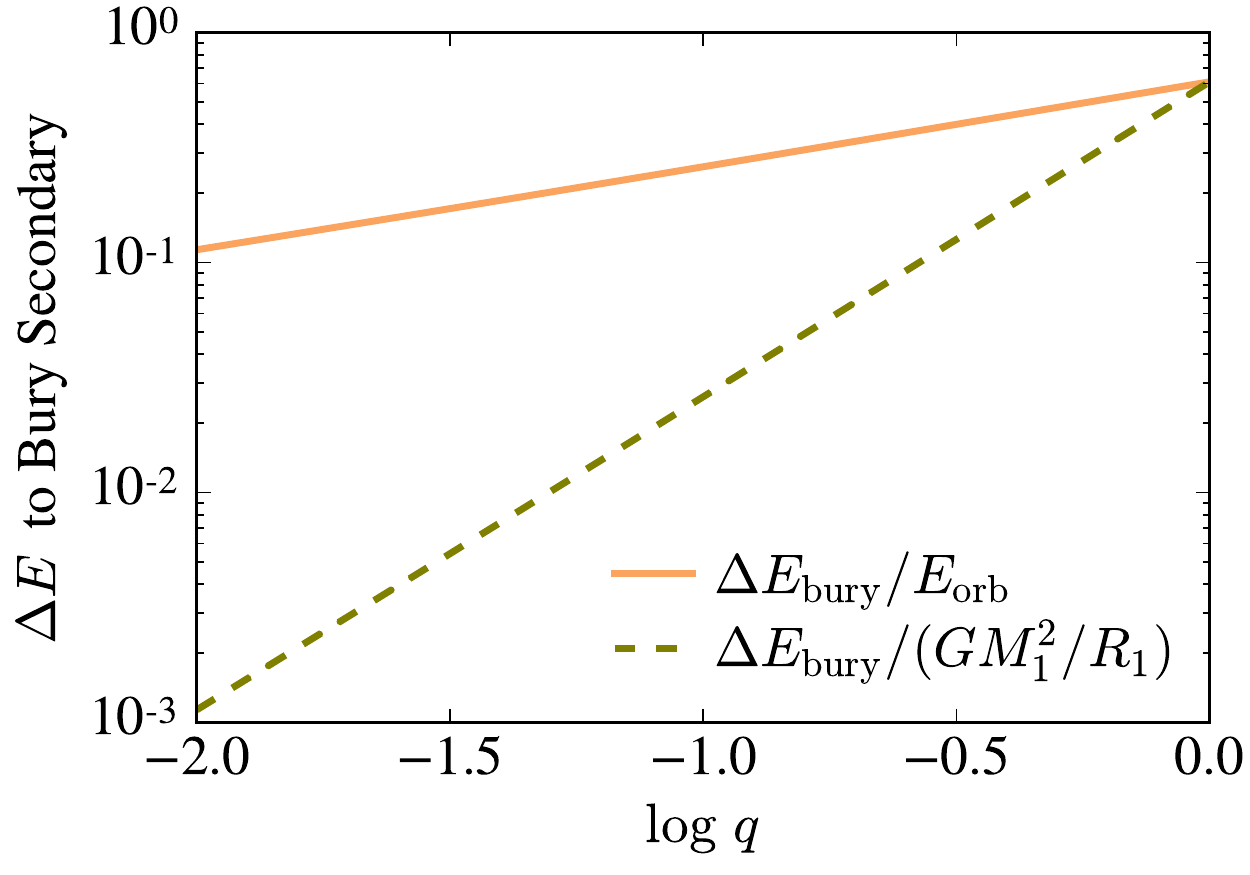}
\caption{ Change in energy to `bury' the secondary object, $\Delta E_{\rm bury}$, defined as the change in orbital energy from $a=R_1$ to $a=R_1 - R_{{\rm RL}_2}$, when the secondary has inspiraled by one Roche radius and is no longer gravitationally dominant for material at the primary's surface. In the figure above, $\Delta E_{\rm bury}$ is compared to the orbital energy at the onset of common envelope, $E_{\rm orb}$, and to the order of magnitude binding energy of the primary, $G M_1^2/R_1$.  The change in orbital energy to bury the secondary is always a fraction of the orbital energy itself, but it also is an even smaller  fraction of the primary's binding energy (more steeply varying with $q$).  }
\label{fig:bury}
\end{center}
\end{figure}

We can place a lower limit on the mass of the secondary object based on the outburst's energetics. In particular, if we assert that the primary source of energy in the transient to be the dissipation of the orbital energy of the secondary into the ejected envelope gas, then we can place a lower limit on the change in orbital energy during this phase (and in turn on $M_2$).\footnote{We note that orbital energy isn't the only possible energy source in a common envelope episode \citep{1993PASP..105.1373I}. The possibility of accretion energy mediated by jets has also been proposed to explain red transients \citep{2016RAA....16...99K, 2016MNRAS.462..217S}.  }
The first peak of the transient outburst is particularly useful in this regard because we have argued that it originates from the phase when the secondary is grazing the surface of the primary. 

The rapid velocities, $v_{\rm ej} \gtrsim v_{\rm esc}$, measured during the outburst peak imply that shocked ejecta are free to expand uninhibited and without thermalizing on the surrounding envelope gas. This suggests a phase of interaction before the secondary object has become `buried' within the envelope of the primary.  In Figure \ref{fig:bury} we plot a quantity $\Delta E_{\rm bury}$, which we define as the magnitude of the change in orbital energy liberated before the secondary is subsumed within the envelope of the primary. 
This energy source must be sufficiently large to explain the early light curve energetics. 
Below, we explore the limits this places on the range of possible mass ratios.

To approximate $\Delta E_{\rm bury}$, we assume that the secondary can eject material to infinity when it is within one (secondary) Roche lobe radius of the surface of the primary. We therefore compute the liberated energy as (minus) the difference in orbital energy  between $a=R_1$ and $a=R_1 - R_{{\rm RL}_2}$,
\beq
\Delta E_{\rm bury} = -\left[E_{\rm orb} \left(  R_1 -R_{{\rm RL}_2} \right) - E_{\rm orb} \left(  R_1  \right) \right],
\eeq
 where  $E_{\rm orb} = - G M_1 M_2 / (2 a)$, and $R_{{\rm RL}_2}$ is the radius of the Roche lobe of the secondary, approximated by the inversion of Eggelton's formula, equation \eqref{RocheLimit} (thus $\Delta E_{\rm bury}>0$ because the orbital energy becomes more negative).  For our definition of $q=M_2/M_1$ we have,
\beq
{R_{{\rm RL}_2} \over a } = \frac{0.49 q^{2/3}} { 0.6q^{2/3} + \ln \left( 1+q^{1/3} \right)  }.
\eeq
In Figure \ref{fig:bury}, we don't include the (few percent) change in enclosed primary mass as the secondary plunges into the outer layers of the primary star's envelope. 
This approximation of $\Delta E_{\rm bury}$ has desirable qualitative features. When the secondary mass is small, it need not plunge deep within the envelope of the primary before the material in its vicinity is thermalized rather than directly ejected. When the secondary mass is large, it instead causes a more global disturbance to the primary envelope and can eject material to infinity from deeper depths within the envelope (smaller $a$).

Returning now to Figure \ref{fig:bury}, $\Delta E_{\rm bury}$ is plotted normalized to the orbital energy at the onset of common envelope ($E_{\rm orb}$ at $a=R_1$) and to the approximate binding energy of the primary, $GM_1^2 / R_1$. The orbital energy released at the onset of common envelope is a steep function of the secondary mass  (roughly $\propto q^{1.36}$) because of the mass dependence not only of the orbital energy at the onset of common envelope, but also of the  depth to which the secondary can penetrate before we assume it to be buried.  We use this information to place a limit on the binary mass ratio by requiring that the liberated energy is larger than the radiated energy of the peak: $\Delta E_{\rm bury}(q) > E_{\rm rad,peak}$, where the radiated energy is plotted as a function of reddening in Figure \ref{fig:erad}. We note that the kinetic energy of the outflow must be larger than the radiated energy (see equation \eqref{ek} and the following discussion). But, because the radiated energy is the observed quantity, this is a strictest constraint that we must ensure is satisfied by any potential system.

\subsubsection{Range of Possible Secondary Masses}

\begin{figure}[tbp]
\begin{center}
\includegraphics[width=0.48\textwidth]{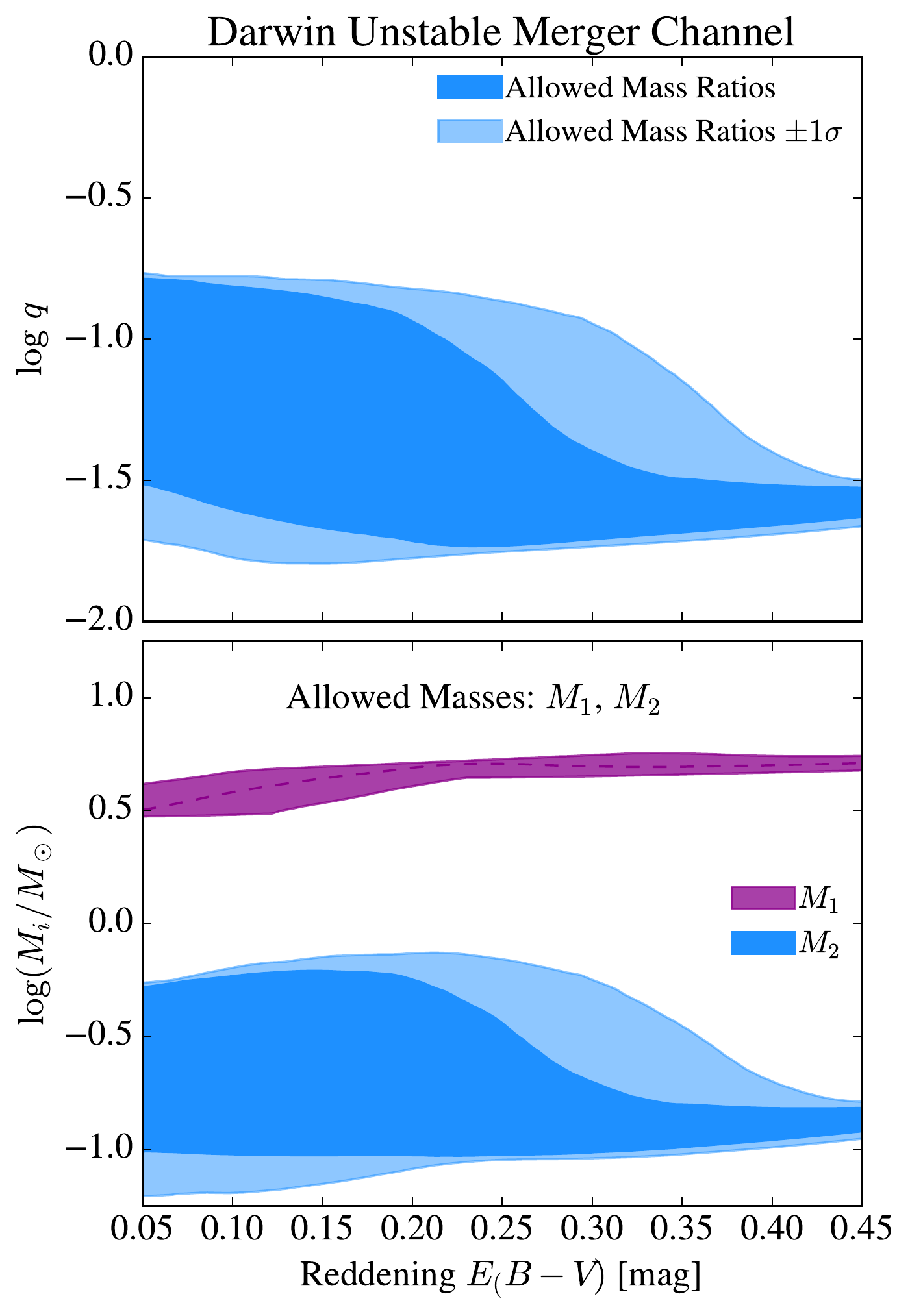}
\caption{ The range of allowed mass ratios ($q$, top panel) and masses ($M_1$, $M_2$, bottom panel)  for M31LRN 2015 under the assumption of a Darwin unstable merger channel. These constraints are derived from the radiated energy on the lower boundary and the Darwin instability on the upper boundary. The lower bound requires that there is sufficient energy to drive the early peak of the observed light curve. The upper bound requires that the binary desynchronize through the Darwin instability. The shape of the upper bound is set by changing internal structure of the inferred primary stars -- which have convective envelopes at low reddening and radiative envelopes at high reddening (see Figure \ref{fig:prog_prop}). In the lower panel, the purple dashed line and shaded region show the inferred constraints on $M_1$, while the blue region uses the constraint on $q$ to deduce $M_2$.    }
\label{fig:allowedq}
\end{center}
\end{figure}

Figure \ref{fig:allowedq} shows the range of mass ratios (top panel) and masses (bottom panel) permitted by the upper and lower limits described above. Since our knowledge surrounding the processes regulating stellar mergers remains quite uncertain, these limits are most appropriately interpreted in terms of the qualitative guidance they offer, rather than an exact numerical value. 

The range of allowed $q$, like the other properties of the binary system,  is a function of the reddening along the line of sight to the source. The upper mass limit, derived on the basis of the Darwin instability, contains the imprint of the internal structure of the primary star, in particular its specific moment of inertia, $\eta_1$.  At low reddening, the primary star properties we derive point to an object with a convective envelope, while at higher reddening, the primary star would be more massive and have a radiative envelope. 

Across all reddening values, the secondary is required to be much less massive than the primary, and $q \ll 1$.  Taking a reddening value of $\ebv = 0.12$~mag, as assumed by \citet{2015ApJ...805L..18W}, this implies that we are observing the merger of a primary star with mass $M_1 \sim 4 M_\odot$, radius of $R_1 \sim 35 R_\odot$ with a low-mass secondary star with mass $M_2 \sim 0.1 - 0.6 M_\odot$.  These quantities (taking the upper-end of the secondary mass range) suggest a pre-merger binary period of $\sim 11$~d. This suggested mass ratio compliments the recent discovery of eclipsing binaries with B-type main sequence stars and extreme mass ratios by \citet{2015ApJ...801..113M,2015ApJ...810...61M}\footnote{For example, \citet{2015ApJ...801..113M} find that 2\% of B-type main sequence stars have low-mass companions in the 3-8.5 day period range similar to the M31LRN 2015 system's binary period.} and the inference of a preference among B-type stars for low-mass companions by \citet{2016AJ....152...40G}.

\section{Discussion and Conclusion}\label{sec:discussion}

\subsection{Summary}

This paper has studied initial photometric and spectroscopic data published by \citet{2015ApJ...805L..18W} and \citet{2015A&A...578L..10K} on a particularly remarkable addition to the growing class of stellar merger-transients. 
This event is especially useful in teaching us about the physics of this process because of the additional detection of a pre-outburst source (with both color, magnitude, and known distance). In this paper we've linked the observational data to modeling of the light curve and pre-outburst source to map the properties of the binary before its merger to the transient it created. 

Our analysis of the M31LRN 2015 transient reveals some likely properties of a binary system at the onset of a common envelope episode.
 A several solar mass star ($\sim4-5.5M_\odot$) evolved away from the main sequence and grew to a radius of $\sim 35 R_\odot$ before engulfing its companion.  The light curve and spectra reveal multiple components: 1) a fast early outflow that becomes transparent at $\sim 2$~AU with characteristic temperature of around $10^4$~K, and 2) a cooler, longer lived outflow which drives the expansion of the photosphere to $\sim 10$~AU, where the temperature is low enough for hydrogen to recombine (and dramatically lower the opacity).  We infer that this early component (with characteristic length similar to one binary orbital period) is driven by shocks that race through the stellar atmosphere during the violent onset of the merger, driving off <1\% of the system mass ($\sim 0.01 M_\odot$) with characteristic velocity similar to the system escape velocity ($v_{\rm ej} \sim 1.5-2 v_{\rm esc}$).  The later, more mass-rich ejecta ($\sim 0.3 M_\odot$) likely arise from the continued inspiral of the secondary after it has plunged deeper and become embedded in the common envelope. 
Our results are broadly consistent with earlier derivations using similar methods of a $\sim 2-4 M_\odot$ progenitor \citep{2015ATel.7173....1D}, and of order $0.1M_\odot$ ejecta mass \citep{2015ApJ...805L..18W}.

One of the most striking features of the M31 LRN 2015 transient is an initial rise time (of the early peak of the lightcurve) of the same order as the inferred binary orbital period. An open puzzle for binary evolutionary scenarios is understanding what drives this rapid transition. Precursor emission, as observed by \citep{2015ATel.7173....1D}, might offer some evidence toward unwrapping the puzzle. 
We attempt to make a step toward answering aspects of this question by extending our analysis to consider the possible pathways to merger for a binary system like that observed in M31LRN 2015 in Section \ref{sec:mergers}.  

We argue in Section \ref{sec:DI} that a scenario in which the binary desychronized and was driven toward merger by the Darwin instability seems like a plausible explanation for the high observed ejecta velocities relative to the system escape velocity. We draw on this possible constraint to work through the implications for the mass of the unseen companion. This analysis shows that, were the system driven to merger by Darwin instability, the binary had a low-mass secondary and a relatively asymmetric mass ratio, $q\sim 0.1$, as shown in Figure \ref{fig:allowedq}, a conclusion which appears compatible with recent analysis of the companion masses of B stars \citep{2015ApJ...801..113M,2015ApJ...810...61M,2016AJ....152...40G}.

\subsection{Future Prospects}

The discovery of M31LRN 2015 marks a step forward in our understanding of flows at the onset of a common envelope episode. 
Although the details of flows in common envelope have remained a theoretical mystery for the past forty years, with observations of common envelope events ``in action''  we are now well positioned to start to constrain this important, yet highly uncertain process in binary evolution \citep{2013Sci...339..433I}.  Indeed, the stakes have never been higher: common envelope episodes are thought to be essential in tightening the orbits of compact binaries to the point that they can merge by gravitational radiation \citep[e.g.][]{2007PhR...442...75K,2014LRR....17....3P} -- as in the recent LIGO detection of merging black holes with massive-star progenitors that likely evolved through a common envelope phase that tightened their orbit \citep{2016ApJ...818L..22A}.  Whether a particular interaction leads to merger or to envelope ejection, is a critical question for forming gravitational wave sources, especially  because the tightest binaries that form are those that are driven {\it nearly} to merger. A more detailed understanding of common envelope ejection is also crucial for resolving uncertainties in the formation of low mass X-ray binaries with either neutron star or black hole primaries. 

There remains much to be learned from other nearby extragalactic transients in which we are able to identify a pre-outburst source. The event rates of these (and similar) events are also promising: \citet{2014MNRAS.443.1319K}  predict galactic rates of $\sim0.1$~yr$^{-1}$ for LRNe, with a steeply-rising number of shallower-amplitude outbursts. The discovery (and analysis) of a similar transient in M101 with extensive data showing a slightly more massive progenitor strengthens this conclusion, and compliments our analysis of M31 LRN 2015 \citep{2016arXiv160708248B,2016AstBu..71...82G}.  It therefore appears reasonable to expect one or more common envelope or stellar merger transients per year in local galaxies with extensive multicolor imaging to rely on for pre-outburst detections.

While this class of objects can teach us many lessons about dynamical phases of binary evolution, there remains much to be learned from M31LRN 2015 itself.  This paper has presented analysis and possible interpretations of the early optical light curve and pre-outburst source. 
Ongoing observations of this object at infrared wavelengths can map out the now-dusty ejecta's mass and energetics. 
The contracting photosphere in the optical observations $\sim50$~d after peak suggests that the binary merged completely and promptly, but future multi-wavelength observations of the remnant object can better constrain the fate of this common envelope episode. Will we observe the remnant's thermal relaxation in future years? Or does a central binary remain enshrouded within the envelope that will instead continue to drive dusty ejecta? 
And, finally, will the spectrum of the merged giant distinguish itself from that of the remnant of the seemingly-similar V838 Mon or from isolated giants of the similar mass and temperature? 
The answers to these questions will be especially interesting in comparison to the properties of the seemingly-similar M31 RV red transient, thirty years further into its post-outburst phase \citep{2011ApJ...737...17B}. 

\begin{acknowledgements}
We acknowledge many useful discussions that lead us to appreciate the extent of M31LRN 2015's potential lessons. 
We are particularly grateful to Suvi Gezari for questions that lead us to pursue this object in detail and for comments on earlier versions of this manuscript. 
We thank M. Darnley, A. Kurtenkov, and S. Williams for correspondence and assistance that facilitated our interpretation of their published data. 
We are grateful for detailed feedback and suggestions from an anonymous referee. 
We also thank 
A. Antoni, F. Antonini, K. Auchettl,  J. Choi,  M. Drout,  M. Fumagalli,  Z. Jennings, K. Kashiyama, 
C. Kochanek,  A. Murguia-Berthier, S. Naoz, O. Pejcha, T. Thompson, M. Trenti, M. Rees, and J. Samsing for discussions, assistance, comments, and ideas. 

This research made use of {\tt astropy}, a community developed
core Python package for Astronomy \citep{2013A&A...558A..33A}. The calculations for this research were carried out in part the UCSC supercomputer Hyades, which is supported by National Science Foundation (award number AST-1229745) and UCSC.
MM acknowledges support from the Chancellor's Dissertation-Year Fellowship at UCSC. 
PM is supported by an NSF Graduate Research Fellowship and a Eugene Cota-Robles Graduate Fellowship. 
ER-R acknowledges financial support from the Packard Foundation, Radcliffe Institute for Advanced Study and NASA ATP grant NNX14AH37G. 
AB is supported by a UC-Mexus Posdoctoral Fellowship. 
GM acknowledges the AAUW American Fellowship. Additional support for this work was provided by NASA through Einstein Postdoctoral Fellowship grant number PF6-170155 awarded by the Chandra X-ray Center, which is operated by the Smithsonian Astrophysical Observatory for NASA under contract NAS8-03060.

\end{acknowledgements}

\bibliographystyle{aasjournal}
\bibliography{merger_refs}

\begin{appendix}

\section{Plateau Scalings In Recombination Transients}\label{sec:recomb}

The plateau portion of red novae's lightcurves can be used to reveal useful properties of the outflow and mass ejection site. Our discussion in this appendix is based on \citet{2013Sci...339..433I}'s application of the analytic theory of recombination transients \citep[e.g.][]{1993ApJ...414..712P,2009ApJ...703.2205K,2010ApJ...714..155K} to luminous red novae transients. 
Hydrogen-rich gas is assumed to be given some energy divided between thermal and kinetic. The gas expands, cooling adiabatically, until it reaches the temperature where hydrogen starts to recombine. This changes the opacity dramatically, and the photosphere settles to the location of the recombination wave within the expanding ejecta and the transient radiates away the adiabatically-degraded internal energy of the ejecta.\footnote{There are assumptions in these models that may complicate their application to merger transients. The internal energy is assumed to decay with expansion $\propto r^{-1}$, which is relevant for radiation-pressure dominated gas with no additional heating or cooling (gas pressure-dominated ejecta would cool as $\propto r^{-2}$). In a stellar merger, \citet{2013Sci...339..433I} has convincingly argued that  hydrogen and helium recombination energies may both contribute significantly to the internal energy budget. If the ejecta has multiple velocity components (as for example has been inferred in classical novae), internal shocks would act as another heating mechanism. Both of these possibilities would contribute to a shallower decay of internal energy with ejecta expansion than the equation of state scalings above might indicate. }  
The luminosity $(L_{\rm p})$ and timescale $(t_{\rm p})$ of the plateau phase are related to the properties of the ejecta in these models by
\beq\label{lp}
L_{\rm p} \approx 4.2 \times 10^{37}~{\rm erg~s}^{-1}  
 \left(R_{\rm init} \over 10 R_\odot \right)^{2/3}
  \left( \Delta M \over 0.1 M_\odot \right)^{1/3}
   \left(v_{\rm ej} \over 100~{\rm km~s}^{-1} \right)^{5/3}
    \left( \kappa \over 0.32~{\rm cm^2~g}^{-1}\right)^{-1/3} 
    \left(T_{\rm rec} \over 4500~{\rm K} \right)^{4/3},
\eeq
and 
\beq\label{tp}
t_{\rm p} \approx  42~{\rm d} \left(R_{\rm init} \over 10 R_\odot \right)^{1/6}
 \left( \Delta M \over 0.1 M_\odot \right)^{1/3} 
 \left(v_{\rm ej} \over 100~{\rm km~s}^{-1} \right)^{-1/3} 
 \left( \kappa \over 0.32~{\rm cm^2~g}^{-1}\right)^{1/6} 
 \left(T_{\rm rec} \over 4500~{\rm K} \right)^{-2/3},
\eeq
where $R_{\rm init}$ is the ejection radius of the gas (perhaps the primary-star radius), $\Delta M$ is the ejecta mass, $v_{\rm ej}$ is the ejecta velocity, $\kappa$ is the opacity of the ionized gas prior to the passage of the recombination wave, and $T_{\rm rec}$ is the recombination temperature of the gas (similar to the photosphere effective temperature to within a factor of $2^{1/4}$ in the grey atmosphere approximation). 
The above expressions (and variables) are identical to Equations (1) and (2) of \citet{2013Sci...339..433I}, with the exception we have substituted in the ejecta velocity rather than the kinetic energy in the original expressions \citep[for more background on these expressions and the underlying model, see section 2 of the supplementary material of][]{2013Sci...339..433I}. 
The total radiated energy during the plateau is, $E_{\rm rad,p} \sim L_{\rm p} t_{\rm p}$, and therefore scales as
\beq\label{Eradp} 
E_{\rm rad,p}\approx 1.5\times 10^{44} \rm{erg}
\left(R_{\rm init} \over 10 R_\odot \right)^{5/6}
 \left( \Delta M \over 0.1 M_\odot \right)^{2/3} 
 \left(v_{\rm ej} \over 100~{\rm km~s}^{-1} \right)^{4/3} 
 \left( \kappa \over 0.32~{\rm cm^2~g}^{-1}\right)^{-1/6} 
 \left(T_{\rm rec} \over 4500~{\rm K} \right)^{2/3},
\eeq
for the same parameters as  \eqref{lp} and \eqref{tp}.

Equations  \eqref{lp} and \eqref{tp} define a system with seven variables: $L_{\rm p}, t_{\rm p}, R_{\rm init},  \Delta M, v_{\rm ej}, \kappa, T_{\rm rec}$. Considering this phase space, 
\citet{2013Sci...339..433I} emphasize that the ejecta velocity might typically be related to the ejecta mass and initial radius if a fraction of a star's mass is assumed to be ejected at the escape velocity, reducing the full parameter variation that might otherwise seem to be present. 
For observed transients, however, we need not make that assumption because we can measure   many of these properties. Taking M31 LRN 2015 as a specific example:
\begin{itemize}
\item The luminosity and duration of the plateau are measured in the lightcurve. 
\item The ejecta velocity can be inferred from spectra and from the photospheric expansion. 
\item The observed effective temperature during the recombination plateau provides an estimate of the recombination temperature. 
\end{itemize}
This leaves three parameters: $\kappa,R_{\rm init},  \Delta M$. We therefore need an estimate of $\kappa$, here we scale to the Thompson opacity in solar composition material. The remaining parameters are astrophysically interesting because they tell us about the radius of mass ejection, and therefore the approximate size of the binary orbit at the time of coalescence and the amount of mass ejected. From \eqref{lp} and \eqref{tp} we can find,
\beq\label{dmp_obs}
\Delta M \approx 0.04 M_\odot 
 \left(v_{\rm ej} \over 100~{\rm km~s}^{-1} \right)^{3} 
 \left( \kappa \over 0.32~{\rm cm^2~g}^{-1}\right)^{-1} 
 \left(T_{\rm rec} \over 4500~{\rm K} \right)^{4}
 \left(t_{\rm p} \over 40~{\rm d} \right)^4
 \left(L_{\rm p} \over 10^{38}~{\rm erg~s}^{-1} \right)^{-1},
\eeq
for the ejecta mass in terms of observable parameters, and 
\beq\label{rip_obs}
R_{\rm init} \approx 61 R_\odot
 \left(v_{\rm ej} \over 100~{\rm km~s}^{-1} \right)^{-4} 
 \left( \kappa \over 0.32~{\rm cm^2~g}^{-1}\right) 
 \left(T_{\rm rec} \over 4500~{\rm K} \right)^{-4}
 \left(t_{\rm p} \over 40~{\rm d} \right)^{-2}
 \left(L_{\rm p} \over 10^{38}~{\rm erg~s}^{-1} \right)^{2},
\eeq
for the ejection radius. Although the right-hand-side parameters are all potentially observable, the utility of these expressions may be limited by the fact that many of the parameters enter with relatively high powers.  
Other combinations of parameters might prove useful in other cases. For example, 
\beq\label{dmp_r}
\Delta M \approx 0.09 M_\odot 
\left(R_{\rm init} \over 10 R_\odot \right)^{-1/2}
 \left(v_{\rm ej} \over 100~{\rm km~s}^{-1} \right)
 \left( \kappa \over 0.32~{\rm cm^2~g}^{-1}\right)^{-1/2} 
 \left(T_{\rm rec} \over 4500~{\rm K} \right)^{2}
 \left(t_{\rm p} \over 40~{\rm d} \right)^3,
\eeq
could be a useful ejecta mass estimator if a measurement or estimate of the progenitor radius was obtained, but the distance (and therefore $L_{\rm p}$) is not known), with the advantage of a weak power in some of the potentially more uncertain parameters like $R_{\rm init}$. In such a scenario one could imagine returning this estimate to equation \eqref{lp} and using the observed flux and estimated plateau luminosity to obtain a rough distance estimate to the source.

\section{Mass Lost from $L_2$ Prior to the Onset of Common Envelope in Roche Lobe Overflow Mergers}\label{sec:dml2}

One pathway to the onset of common envelope is mass loss from the outer, $L_2$, Lagrange point. This material forms a cool, equatorial outflow with characteristic radial velocity significantly less than the system escape velocity \citep{1979ApJ...229..223S,2016MNRAS.455.4351P,2016MNRAS.461.2527P}. 

We can calculate how much mass would need to be lost to carry away the angular momentum needed to bring the binary from $a=a_{L_2}$ to the onset of common envelope at $a=R_1$.  The orbital angular momentum of the binary is 
\beq\label{Lorb}
L_{\rm orb} = \mu \left( G M a \right)^{1/2}.
\eeq
  Thus the change in orbital angular momentum between $a=a_{L_2}$ and $a=R_1$(for the moment neglecting the loss in mass) is 
\beq
\Delta L_{\rm orb} \approx L_{\rm orb} \left(a_{L_2} \right) - L_{\rm orb} \left(R_1 \right). 
\eeq
Mass loss from the $L_2$ point carries angular momentum away from the binary at a rate of
\beq\label{dLdm}
{{\rm d} L \over {\rm d} m} = \left( G M a \right)^{1/2} r_{L_2}^2,
\eeq
where $r_{L_2} = x_{L_2}/a$ is a dimensionless ratio, dependent on $q$, which describes the distance to the $L_2$ point from the center of mass of the binary in units of the semi-major axis \citep{1998CoSka..28..101P}. Typical values for $r_{L_2} \approx 1.2$ and it varies by <10\% across a wide range of $q$.  We can then estimate a mass loss $\Delta m_{L_2} \approx \Delta L_{\rm orb} / ({\rm d}L/{\rm d}m)$, where, for simplicity, we evaluate ${\rm d}L/{\rm d}m$ at $a_{\rm L_2}$. This gives,
\beq\label{dm2_simple}
{\Delta m_{L_2} \over M_1} \approx {q\over 1+q} r_{L_2}^{-2} \left[  1- \left( {R_1 \over a_{L_2}} \right)^{1/2} \right]. 
\eeq
We've plotted this quantity in the right-hand panels of Figure \ref{fig:sep} labeled as `Simple'. 

To obtain a more numerically accurate version of $\Delta m_{L_2}$, we need to integrate the mass lost as the binary separation decreases from $a=a_{L_2}$ to $a=R_1$. We use equations \eqref{dLdm} and \eqref{Lorb} to express 
${\rm d}m/{\rm d}a = ({\rm d}m/{\rm d}L) ({\rm d}L/{\rm d}a)$, to find
\beq\label{dmda}
{{\rm d}m \over {\rm d}a} = - {\mu \over 2 a }r_{L_2}^{-2}.
\eeq
We integrate ${\rm d}m/{\rm d}a$ numerically from $a=a_{L_2}$ to $a=R_1$ to find $\Delta m_{L_2}$. We use the midpoint method and $10^3$ evenly-spaced integration steps. We assume that mass lost comes from the envelope of the primary, thus decreasing $M_1$ and modifying $\mu$ and $q$ through the integration.  This numerically derived version of $\Delta m_{L_2}$ is slightly higher than the estimate of equation \eqref{dm2_simple} and is plotted in the right-hand panels of Figure \ref{fig:sep} labeled `Numerical'. 

Figure \ref{fig:sep} shows that the mass loss needed to bring the binary from the point where it begins to shed material to the onset of common envelope is of order 10\% of $M_2$. 
This calculation makes the assumption that Roche lobe overflow occurs when the binary reaches $a=a_{L_1}$. The desynchronization of the orbital motion and the primary's rotation can substantially suppress mass transfer in systems that first destabilize through the Darwin instability. \citet{2007ApJ...660.1624S} have shown that non-synchronous systems need to reach closer separations before the effective potential allows material to flow from one Roche lobe to another. In these systems the initial mass loss might instead come when material is ejected at the onset of common envelope, not through Roche Lobe overflow. 

\begin{figure}[tbp]
\begin{center}
\includegraphics[width=0.75\textwidth]{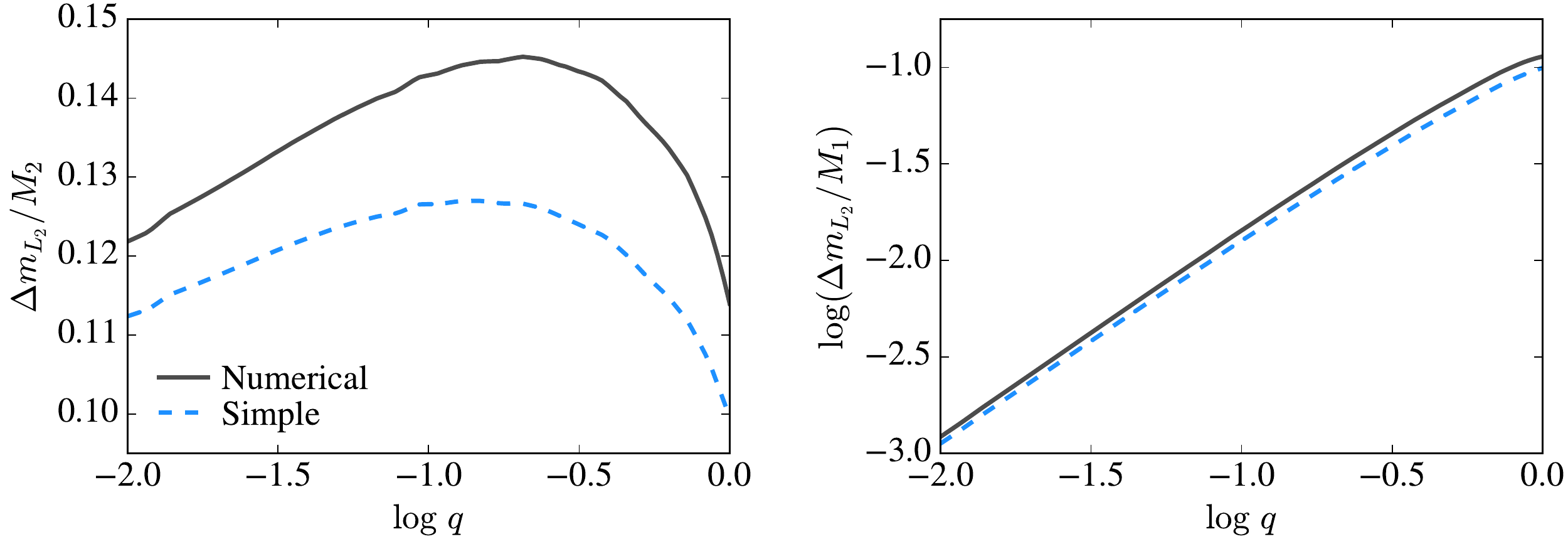}
\caption{Mass lost from the outer Lagrange point, $L_2$, prior to the onset of common envelope in systems merging by Roche lobe overflow. This is the mass loss needed to carry away the orbital angular momentum of the binary. Two estimates of this mass loss are shown, the first, labeled ``Simple'' is based on equation \eqref{dm2_simple}, while the second, which is derived from a numerical integration of equation \eqref{dmda}, is labeled ``Numerical". Approximately 10-15\% of $M_2$ is needed to drive systems to merger through Roche lobe overflow. This material is trailed off from the $L_2$ point and expands with radial velocity $\lesssim 0.25 v_{\rm esc}$ \citep{1979ApJ...229..223S,2016MNRAS.455.4351P,2016MNRAS.461.2527P}. This substantial mass loss shapes the circumbinary environment. The pile-up of slow-moving material might also decelerate less massive but faster shock-driven ejecta from the onset of common envelope.  }
\label{fig:dml2}
\end{center}
\end{figure}

\section{Comparison to Other Stellar-Merger Transients}

The similarities of the optical light curve and spectra of M31LRN 2015 to transients like V838 Mon enabled its early identification as a stellar merger and triggered the extensive follow-up observations published by \citet{2015ApJ...805L..18W} and \citet{2015A&A...578L..10K}.  The class of binary merger transients has grown to contain objects with a range of properties. In this section we briefly compare of M31LRN 2015 to several recent and well-studied transients which may share similar origin: V1309 Sco, V838 Mon, M101OT2015-1, and NGC4490-OT.  

V1309 Sco was a particularly useful detection in defining the class of stellar merger transients because an eclipsing binary with a decreasing orbital period was observed by the OGLE  experiment \citep{2008AcA....58...69U} for nearly a decade prior to the event \citep{2011A&A...528A.114T}.  The eclipsing binary has an inferred mass ratio of $q\approx 0.1$, \citep{2016RAA....16...68Z}. Eventually, the modulation of the light curve disappeared and was replaced by a short dimming then steady brightening over approximately 6 months. \citet{2014ApJ...788...22P} have associated this phase with mass loss from the $L_2$ Lagrange point, which is carrying away orbital angular momentum and driving the binary toward merger as it also obscures the central objects and leads to a growing (and brightening, due to internal shocks in the mass loss) photosphere.  This phase of steady brightening transitions smoothly to an abrupt peak. \citet{2014ApJ...786...39N} and \citet{2014ApJ...788...22P} both associate this abrupt peak with shocks driven through the outflow at the onset of common envelope when one star plunges within its companion,  and \citet{2013Sci...339..433I} estimate an ejecta mass $\sim 0.03 M_\odot$.  Many similarities between V1309 Sco's evolution and the qualitative phases of the M31LRN 2015 outburst present themselves. In particular, a fast, shock-driven outflow seems likely to drive the light curve peak. 
However, the characteristic velocities measured from the H$\alpha$ FWHM in V1309 Sco ($ 150$ km s$^{-1}$) are only $\sim 40$\% of  the modeled primary-star's ($M_1 \approx 1.5 M_\odot$, $R_1 \approx 3.5 R_\odot$) escape velocity ($\approx 400$ km s$^{-1}$) \citep{2014ApJ...786...39N}. 
By contrast, for M31LRN 2015, velocities of $v_{\rm ej} \gtrsim v_{\rm esc}$ are observed. In Section \ref{sec:DI}, we argued that one possible explanation for these high relative velocities is a comparatively unpolluted circumbinary environment without substantial $L_2$ mass loss to decelerate the early ejecta -- implying different channels to merger for M31 LRN and V1309 Sco.

V838 Mon's light curve shows an even more dramatic multiple-component structure than either that of M31LRN 2015 or V1309 Sco, with at least three individual peaks \citep[e.g.][]{2003Natur.422..405B}. V838 Mon's merger generated significantly faster outflows than V1309 Sco, with velocity $\sim 500$ km s$^{-1}$ \citep{2002A&A...389L..51M}.  It reaches a very similar peak luminosity to M31LRN 2015 \citep{2015ApJ...805L..18W}.  Light echos from the outburst allowed the determination of an accurate distance \citep{2003Natur.422..405B,2005A&A...434.1107M,2008AJ....135..605S}.
The system contains a distant B-type companion \citep{2007A&A...474..585M,2010ApJ...717..795A}.
However, there remains discussion on the nature of the progenitor, with arguments for a young (main sequence or pre-main sequence) B-type $(5-10 M_\odot)$ primary \citep{2005A&A...441.1099T,2007AJ....133..387A} or a hotter star that was initially substantially more massive  \citep[$\sim65M_\odot$,][]{2005A&A...434.1107M}. 
Recently, \citet{2014A&A...569L...3C} and \citet{2015AJ....149...17L} have found that a decade after its original outburst, V838 Mon exhibits a very cool, extended L-type supergiant remnant. 
The uncertainty surrounding the nature of the progenitor of V838 Mon makes it difficult to draw firm conclusions about this object's evolutionary history. Even so, the light curves show distinctly similar color evolution and duration to that of M31LRN 2015 \citep[as emphasized by ][in their discussion of the light curve]{2015ApJ...805L..18W}, perhaps suggesting similar ejecta masses or hydrodynamics. 

Detailed observations and analysis of an extragalactic luminous red nova in M101  (M101 OT2015-1) by \citet{2016arXiv160708248B} and \citet{2016AstBu..71...82G} serve as an extremely valuable compliment to the data presented by \citet{2015ApJ...805L..18W}, \citet{2015A&A...578L..10K}, and this paper on M31 LRN 2015.  \citet{2016arXiv160708248B} detect the progenitor system of M101 OT2015-1 in multicolor imaging spanning 15 years. These data reveal a $L_1\approx 2.6\times10^5 L_\odot$ progenitor with $T_{\rm eff,1}  \approx 6600$~K, and $R_1\approx 220 R_\odot$ -- implying a mass $\sim 18 M_\odot$ \citep[see Section 3.3 of][for details]{2016arXiv160708248B}. This places the progenitor in a portion of the HR diagram where it is expanding across the Hertzsprung gap. Interestingly, and in parallel with V838 Mon, the transient appears to have two peaks separated by $\sim 100$~d.  Not much data is available for the first peak (the object was behind the Sun) but the second peak is well covered with extensive photometric and spectroscopic data. The bolometric lightcurve, \citet{2016arXiv160708248B}'s Figure 8, shows significant similarity to M31 LRN 2015 with a peak followed by plateau morphology. The photosphere expands (at first peak), recedes,  the expands again toward the second peak and a plateau phase (this second expansion velocity is $\approx 115$~km~s$^{-1}$). The H$\alpha$ line profiles in M101 OT2015-1 are complex and intriguing. The overall profile is broad with FWHM $\sim 500$~km~s$^{-1}$, and the profile also shows evidence for a component blueshifted by $500$~km~s$^{-1}$ which transitions from absorption during the peak to emission during the plateau (\citet{2016arXiv160708248B}'s Figure 6). The comparison of these line profiles to the primary's escape velocity ($\approx 180$~km~s$^{-1}$) reveals high ejecta velocities relative to the system escape velocity, much like M31 LRN 2015. The plateau phase of the bolometric lightcurve has a duration of $\sim 100 - 200$~d. If we apply the scaling of equation \eqref{dmp_r}, we find an ejecta mass of $1-10 M_\odot$, which could indicate that a large fraction of the envelope mass was ejected in this event.  

The properties of NGC 4490-OT were recently presented by \citet{2016MNRAS.458..950S}, who emphasize the similarity of this event in color evolution and light curve structure to other stellar-merger transients -- and its distinction from the supernova 2008S-class intermediate luminosity optical transients. The optical transient is significantly brighter than the others mentioned here, peaking at $M_R\sim-14.2$, and the duration is $\sim 200$ days. The radiated energy during this time is $1.5 \times 10^{48}$~erg, or approximately 2 orders of magnitude larger than that of M31LRN 2015. The scalings of Section \ref{sec:early_peak_scalings} suggest that the ejection of several solar masses would reproduce these timescales and energetics given the ejecta velocities of several hundred km~s$^{-1}$.    Similar considerations (along with spectroscopic similarities to V838 Mon) lead \citet{2016MNRAS.458..950S} to suggest that NGC 4490-OT could be the massive star equivalent of the stellar-merger transients mentioned previously. Indeed, the detection of a progenitor (though single-color) point toward a massive, late B-type star of $\sim 30 M_\odot$. With a major portion of the lightcurve missing while the transient was obscured by the sun it is difficult to estimate an ejecta mass.  However, even if the ejecta mass was a large fraction of the envelope, hydrogen recombination could play a major role in the radiated energetics, as indicated by equation \eqref{dm_plateau} and \citet{2013Sci...339..433I,2015MNRAS.447.2181I}, but the full transient luminosity and radiated energy remain difficult to explain (under simplistic assumptions) by $\sim 1$ order of magnitude.

Some diversity in these events should absolutely be expected. After all, with two (or more) stars in a stellar multiple system, there are innumerable combinations of stellar mass and type that could be achieved at the onset of merger. 
It is interesting to note that the range of masses estimated for progenitor stars involved in driving merger-transients spans a huge range -- from sources with primaries near a solar mass to those with massive stars. \citet{2014MNRAS.443.1319K} have noted from this evidence that the peak luminosity of  merger transients scales steeply with progenitor mass, perhaps exchanging with the stellar initial mass function to make transients similarly observable across a wide range in progenitor masses. If this trend in the data continues to hold, it will allow us to probe the binary evolution of a range of both low-mass and massive stars.

\end{appendix}

\end{document}